\documentclass[aps,prc,showpacs,twocolumn,superscriptaddress,nofootinbib,preprintnumbers,floatfix]{revtex4-1}

\usepackage{longtable}
\usepackage{graphicx,amsmath}
\usepackage{slashed}
\usepackage{dcolumn}
\usepackage{bm}
\usepackage{amsmath,amssymb,mathrsfs}
\usepackage{epstopdf} 
\usepackage{subfigure}
\usepackage{tensor}
\usepackage{gensymb}
\usepackage{float}
\usepackage{graphics}

\allowdisplaybreaks

\newcommand{\bea}{\begin{eqnarray}}
\newcommand{\eea}{\end{eqnarray}}
\newcommand{\beq}{\begin{equation}}
\newcommand{\eeq}{\end{equation}}

\usepackage[normalem]{ulem}  
\usepackage[dvips]{color} 

\renewcommand\sout{\bgroup \color{red} \ULdepth=-.5ex \ULset}

\begin{document}



\title{Low-Energy Lepton-Proton Bremsstrahlung via Effective Field Theory }


\author{Pulak~Talukdar}
\email[]{t.pulak@iitg.ac.in}
\affiliation{Department of Physics, Indian Institute  of Technology Guwahati, 781039 Assam, India} 
\author{Fred~Myhrer} 
\email[]{myhrer@physics.sc.edu}
\affiliation{Department of Physics and Astronomy, University of South Carolina, Columbia, SC 29208, USA} 
\author{Ghanashyam~Meher} 
\email[]{ghanashyam@iitg.ac.in}
\affiliation{Department of Physics, Indian Institute  of Technology Guwahati, 781039 Assam, India} 
\author{Udit Raha}
\email[]{udit.raha@iitg.ac.in}
\affiliation{Department of Physics, Indian Institute  of Technology Guwahati, 781039 Assam, India}


\date{\today}

\begin{abstract}
We present a systematic calculation of the cross section for the lepton-proton bremsstrahlung process 
$l+p\to l^\prime +p+\gamma$ in chiral perturbation theory at next-to-leading order. This process corresponds 
to an undetected background signal for the proposed MUSE experiment at PSI. MUSE is designed to measure elastic 
scattering of low-energy electrons and muons off a proton target in order to extract a precise value of the 
proton's r.m.s. radius. We show that the commonly used {\it peaking approximation}, which is used to evaluate the 
{\it radiative tail} for the elastic cross section, is not applicable for muon-proton scattering at the low-energy 
MUSE kinematics. Furthermore, we point out a certain pathology with the standard chiral power counting scheme 
associated with electron scattering, whereby the next-to-next-to-leading order contribution from the pion loop 
diagrams is kinematically enhanced and numerically of the same magnitude as the next-to-leading order corrections. 
We correct a misprint in a commonly cited review article.   
\end{abstract}




\maketitle

\section{Introduction }
Recent high precision experimental determinations of the proton's r.m.s. radius have produced results which 
are not consistent with earlier results~\cite{pohl10,Mohr2012,pohl13}. The {\it proton radius puzzle} refers to the 
contrasting results obtained between the proton's electric charge radii extracted from the Lamb shift of muonic 
hydrogen atoms and those extracted from electron-proton scattering measurements, [see, e.g., 
Refs.~\cite{Bernauer2010,Zhan2011,Mihovilovic2017}.] We note that a dispersion relation analysis of the proton 
form factor determined from electron-proton scattering data favors a charge radius~\cite{Ulf1,Ulf2} consistent 
with that extracted from the experimental muonic hydrogen result. In order to resolve this issue, a number of 
newly commissioned  experiments are underway, along with proposals of redoing the lepton-proton scattering 
measurements at low momentum transfers. The latter includes the MUon-proton Scattering Experiment (MUSE)~\cite{gilman13}. 
The MUSE collaboration proposes to measure the elastic differential cross sections for $e^\pm p$ and $\mu^\pm p$ 
scattering at very low momentum transfers with an anticipated accuracy of a few tenths of a percent. This should 
allow for a very precise determination of the slope of the proton's electric form factor $G^p_E$, and thereby, 
extract a value for the proton's r.m.s.  charge radius squared, 
$(r^p_E)^2 = \left. 6\frac{\partial G^p_E(\tau^2)}{\partial \tau^2}\right|_{\tau^2=0}$, where $\tau$ is the   
four-momentum transferred to the proton. The MUSE collaboration aims for  an r.m.s. radius uncertainty of about 
0.01 fm. However, in MUSE only the lepton scattering angle $\theta$ is detected. The final scattered lepton energy 
$E_l^\prime$ is not measured, nor are the bremsstrahlung photons. The bremsstrahlung photons constitute an integral 
part of the lepton-proton elastic scattering, and is one of the principal sources of uncertainties in the accurate 
measurement of the momentum transfers. In order to determine the proton charge radius, the data analysis necessarily 
needs to correct for  this radiation process. 

The lepton beam momenta considered by MUSE are  of the order of the muon mass~\cite{gilman13,LinLi}. 
A particular concern is the standard radiative correction procedure, which makes use of the  
so-called  {\it peaking approximation}, [see, e.g., Refs.~\cite{maximon1969,maximon00,tsai1961,motsai1969} 
for reviews].\footnote{Ref.~\cite{maximon1969} nicely explains the distinctions between radiative corrections 
and the radiative bremsstrahlung tail of the elastic process involving lepton scattering.}  
This approximation assumes that the bremsstrahlung photons are emitted either along the incident beam direction, or 
in the direction of the scattered final lepton momentum. The validity of this approximation normally relies on 
elastic scattering of highly relativistic particles, like electrons. This is, however, questionable when either 
the  particle energy is comparable to its mass, e.g., in the case of low-energy muon scattering in MUSE,    
or for inelastic scatterings with large energy losses (30-40\%) from the incident projectiles due to 
bremsstrahlung~\cite{motsai1969}. The main purpose of this paper is to accurately assess the bremsstrahlung 
processes from electron and muon scattering off a proton at low energies in a model independent formalism. We 
show in a pedagogical manner the radical differences between the electron and muon angular bremsstrahlung spectra. 
We also correct a misprint in Eq.~(B.5) of Ref.~\cite{motsai1969}, which is important in the low-energy lepton 
scattering processes. 

At  low energies, which includes the kinematic region for MUSE, hadrons are the relevant degrees of freedom where 
the dynamics are  governed by chiral symmetry requirements.  {\it Heavy Baryon Chiral Perturbation Theory} 
($\chi$PT) is a low-energy hadronic effective field theory (EFT), which incorporates the underlying symmetries 
and symmetry breaking patterns of QCD. In $\chi$PT the evaluation of  observables follow  well-defined chiral 
power counting rules, which  determine the dominant {\it leading order} (LO) contributions, as well as the 
{\it next-to-leading order} (NLO) and higher order corrections to observables in a perturbative scheme. 
For example, in  $\chi$PT the proton's r.m.s. radius enters at {\it next-to-next-to-leading order} (NNLO)
where pion loops at the proton-photon vertex enter into the calculation, see, e.g., Ref.~\cite{bernard1995}. 
We note that $\chi$PT naturally includes the photon-hadron coupling in a gauge invariant way. 

As mentioned, the goal of the present paper is to provide a pedagogical and model independent presentation 
of the bremsstrahlung process  in a low-energy EFT framework. We will show how the lepton mass     
will crucially influence the  photon's angular distribution, and also discuss how the {\it radiative tail} cross 
section d$\sigma$/(d$p^\prime$d$\Omega^\prime)$ depends on the outgoing lepton momentum $p^\prime$ and its mass $m_l$. 
We demonstrate that the so-called {\it peaking approximation} is not applicable for muon scattering at MUSE 
momenta. In fact, our results corroborate the analysis in Ref.~\cite{ent01}, where it was shown that the 
peaking approximation, which is predominantly valid in the zero-mass limit or for very high-momentum transfers,  
becomes questionable at lower energies and could  lead  to significant errors in estimating the low-energy 
radiative cross sections. In this work we present a systematic evaluation the bremsstrahlung process using 
$\chi$PT up to and including NLO in chiral counting. Furthermore, we provide a rough estimate the NNLO contributions 
from the proton's structure effects that arise from pion loop corrections to the LO proton-photon vertex. These NNLO 
pion loop contributions effectively introduce the proton's r.m.s. radius, the first momentum dependent term in 
the charge form factor of the proton. As we shall discuss, contrary to the expectations based on standard chiral 
power counting for this process, the NNLO pion loop contributions appear to be  kinematically enhanced by the 
small electron mass and roughly of the same order as the NLO contributions. This interesting observation, that 
emerges from our analysis in the case of the electron scattering, is not manifested in the case of the muon 
scattering due to the much larger muon mass.\footnote{
By LO, we mean the correction terms that are leading order in chiral counting, which includes the leading 
kinematic recoil corrections [see Sec.~\ref{section:cross-sections} for  clarification.]} 

The paper is organized as follows: In Sec.~\ref{section:EFT} we present a brief description of the 
lepton-proton bremsstrahlung process and the associated kinematics in the context of $\chi$PT, which, 
in principle, allows an order by order perturbative evaluation of the radiative and proton recoil correction 
contributions. In Sec.~\ref{section:cross-sections} we define the coordinate system and discuss the kinematics 
that are used in our evaluations of the analytic expressions for the differential cross sections. In 
Sec.~\ref{section:results_LO+NLO}, we present the results for our systematic evaluation up to and including NLO 
in chiral power counting. Sec.~\ref{section:summary} contains a discussion of our numerical results. We 
furthermore include an estimate of the NNLO pion loop corrections, and present a comparison of these NNLO results 
with our full NLO evaluations. Finally, a short summary is presented before drawing some conclusions. For the sake 
of completeness we have added an appendix containing explicitly some of our elaborate expressions for the NLO 
amplitudes. 

\section{Low-Energy Lepton-Proton Bremsstrahlung}
\label{section:EFT}
In our evaluation the standard relativistic lepton-current~\cite{BjDr} is given by the expression
\beq
     J^\mu_l(Q)=e \bar{u}_l(p^\prime_l )\gamma^\mu u_l(p_l)\,, 
\eeq
where $e=\sqrt{4\pi\alpha}$, and the four-momentum transferred to the proton is $Q=p_l-p^\prime_l$. The lepton mass 
is included in all our expressions, and we will show that the lepton mass plays a crucial role in determining 
the shape of the low-energy lepton-proton bremsstrahlung differential cross section. The hadronic current is 
derived from the $\chi$PT Lagrangian. In $\chi$PT it is assumed that the LO terms give the dominant contributions 
to the amplitude, while the higher chiral orders contribute smaller corrections to the LO amplitude. The effective 
$\chi$PT Lagrangian ${\mathcal L}_\chi$ is expanded in increasing chiral order as  
\beq
      {\mathcal L}_\chi={\mathcal L}^{(0)}_{\pi N}+{\mathcal L}^{(1)}_{\pi N} 
      +{\mathcal L}^{(2)}_{\pi N} +{\mathcal L}^{(0)}_{\pi \pi} +\cdots \; , 
      \label{eq:Lchiral}
\eeq 
where the superscript in ${\mathcal L}^{(\nu)}$ denotes interaction terms of {\it chiral order} $\nu$.\footnote{ 
The {\it chiral order} is given by the relation $\nu = d+\frac{n}{2}-2$, where $d$ is order of derivative and $n$ is the 
number of nucleons at the vertex. In terms of standard momentum power counting, this corresponds to 
${\mathcal O}({\mathscr{P}}^{(\nu+1)})$, where $\mathscr{P}$ denotes the typical four-momentum transfer in a given process.} 
The proton mass $m_p$ is large, of the order of the chiral scale $\Lambda_\chi \sim 4\pi f_\pi \sim 1$ GeV, where $f_\pi=93$ 
MeV is the pion decay constant. An integral part of $\chi$PT is  the expansion of the Lagrangian in powers of $m_p^{-1}$, 
where at LO, i.e., ${\cal L}^{(0)}_{\pi N}$, the proton is assumed static ($m_p\to\infty$). Note that in our evaluation 
the $m_p^{-1}$ corrections to the bremsstrahlung process have two different origin, namely, the {\it kinematic} 
phase-space corrections, as seen in Eqs.~(\ref{eq:sigma1}) and (\ref{eq:Eprimerecoil}) below, and the {\it dynamical} 
$m_p^{-1}$ NLO corrections that arise from the photon-proton interaction in ${\cal L}_{\pi N}^{(1)}$, Eq.(\ref{eq:Lchiral}). 
In particular, it should be mentioned that in general the anomalous magnetic moments of the nucleons formally enter into 
the $\chi$PT calculations at NLO through  ${\cal L}_{\pi N}^{(1)}$. 

\begin{figure}[tbp]
\centering
         \includegraphics[width=\columnwidth,height=3cm]{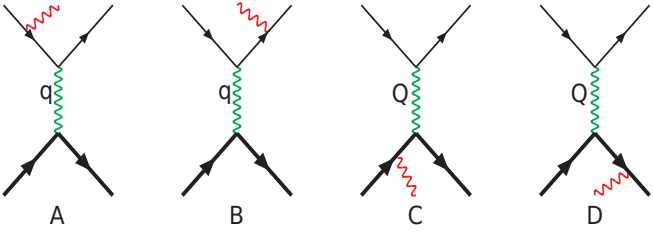}
\caption{Feynman diagrams contributing to lepton-proton bremsstrahlung process. In the Coulomb gauge 
the proton radiation diagrams (C) and (D) do not contribute at the leading order in $\chi$PT.}  
\label{fig:LOfeyndiag}
\end{figure}

We first evaluate the LO contributions to the lepton-proton bremsstrahlung process shown in Fig.~\ref{fig:LOfeyndiag}, 
using the explicit expression for the LO Lagrangian ${\mathcal L}^{(0)}_{\pi N}$ relevant for the processes under study, 
namely~\cite{bernard1995,fettes00},
\bea
      \mathcal{L}^{(0)}_{\pi N}=\bar{N}(iv\cdot D+g_A S\cdot u)N\,\,;\,\, u_\mu &=& iu^\dagger\nabla_\mu U u^\dagger\,.
       \label{eq:Lchiral0}
\eea 
Here $N$ denotes the heavy nucleon spinor field and $g_A=1.26$ is the nucleon axial vector coupling constant, 
which does not contribute to our amplitude at LO and NLO. It is convenient to choose the proton four-velocity 
to be $v^\mu=(1,{\vec{0}})$ which determines the covariant proton spin operator to be $S^\mu=(0,\frac{1}{2}\vec{\sigma})$. 
The covariant derivative $D_\mu$ and $\nabla_\mu$ in Eq.~(\ref{eq:Lchiral0}) are defined as 
\bea
       D_\mu &=& \partial_\mu+\Gamma_\mu-iv^{(s)}_\mu; \nonumber \\
       \Gamma_\mu &=& \frac{1}{2}[u^\dagger(\partial_\mu-ir_\mu)u+u(\partial_\mu-il_\mu)u^\dagger]\,, \\
       \nabla_\mu U &=& \partial_\mu U-ir_\mu U+iUl_\mu\, .
\eea 
The photon field ${\mathcal A}_\mu(x)$ is the only external source in this work, and  
the iso-scalar part of the photon field enters as $v^{(s)}_\mu(x)=-e\frac{I}{2}{\mathcal A}_\mu(x)$,    
where $I$  is the identity SU(2) matrix. 
The {\it chiral connection} $\Gamma_\mu$ and the {\it chiral vielbein} $u_\mu$  include the external 
iso-vector sources $r_\mu(x)$ and $l_\mu(x)$. In our work  the left- and right- handed sources become  
$r_\mu(x)=l_\mu(x)=-e\frac{\tau_3}{2}{\mathcal A}_\mu(x)$. Generally, the $U$-field depends non-linearly on 
the pion field ${\boldsymbol\pi}$, and in the so-called ``sigma''-gauge has the following form: 
$U(x) \equiv u^2(x)=\sqrt{1-{\boldsymbol\pi}^2/f_\pi^2}+i{\boldsymbol\tau}\cdot {\boldsymbol \pi}/f_\pi$. 
It may be noted that $U(x) = 1$ at the LO where explicit pion fields are absent. The pion fields in $U(x)$ 
only enters the calculation to generate the NNLO pion loop contributions, see later. We adopt the Coulomb 
gauge, i.e., $\epsilon\cdot v=0$, where $\epsilon^\mu$ is the outgoing photon polarization four-vector. This 
implies that in $\chi$PT the bremsstrahlung photon from the proton, diagrams (C) and (D) in 
Fig.~\ref{fig:LOfeyndiag}, do not contribute to the bremsstrahlung process at LO. 
%
%

\begin{figure}[b]
\centering
         \includegraphics[width=\columnwidth]{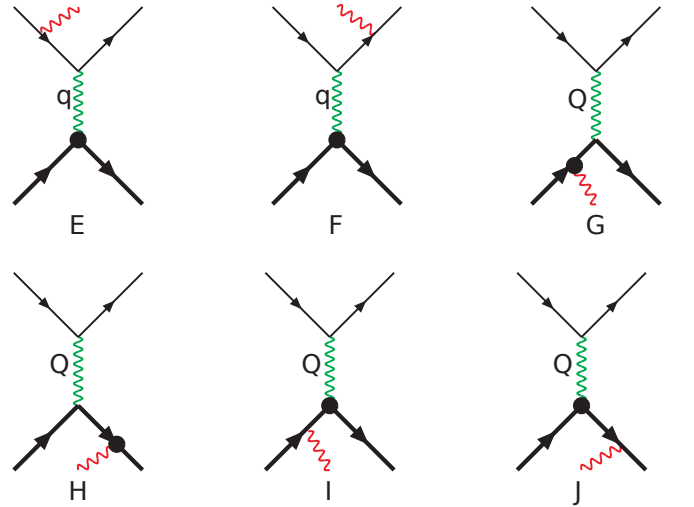}
\caption{Feynman diagrams contributing to the lepton-proton bremsstrahlung process at NLO. The filled 
blobs represents insertion of proton-photon interaction terms from ${\cal L}_{\pi N}^{(1)}$. In the Coulomb 
gauge the proton radiation diagrams (I) and (J) do not contribute. 
}
\label{fig:NLOfeyndiag}
\end{figure}

The first non-trivial contribution of photon radiation from proton [Feynman diagrams (G) and (H) in 
Fig.~\ref{fig:NLOfeyndiag}] arises from the NLO interactions specified by ${\cal L}^{(1)}_{\pi N}$ in 
Eq.(\ref{eq:Lchiral}). Like diagrams (C) and (D) in Fig.~\ref{fig:LOfeyndiag},  diagrams (I) and (J) 
in Fig.~\ref{fig:NLOfeyndiag} do not contribute in the Coulomb gauge. For completeness, we also specify   
the NLO Lagrangian~\cite{bernard1995,fettes00}, where, however, only the terms directly relevant to our 
analysis are retained. 
\bea
{\cal L}^{(1)}_{\pi N} &=& \bar{N}\frac{1}{2m_p}\biggl\{(v\cdot D)^2-D\cdot D -\frac{i}{2}\left[S^\mu,S^\nu\right] \nonumber \\
&& \times \biggl[(1+\kappa_v)f^{+}_{\mu\nu} +2(1+\kappa_s)v^{(s)}_{\mu\nu}\biggr]+\cdots\biggr\}N\,. 
\label{eq:Lchiral1}
\eea
Here 
\bea
f^+_{\mu\nu}&=&u^\dagger F^R_{\mu\nu}u +uF^L_{\mu\nu}u^\dagger \nonumber \\
&=&-e\left(\partial_\mu {\mathcal A}_\nu-\partial_\nu {\mathcal A}_\mu\right)(u^\dagger {\mathcal Q}u +u{\mathcal Q}u^\dagger), 
\label{eq:fmunu}
\eea
where ${\mathcal Q}={\rm diag}(1,0)$ is the nucleon charge matrix, and
\bea
F^{L}_{\mu\nu}&=& F^{R}_{\mu\nu} = \partial_\mu l_\nu-\partial_\nu l_\mu-i\left[l_\mu,l_\nu\right] \nonumber \\
&=&-e\frac{\tau_3}{2}\left(\partial_\mu {\mathcal A}_\nu-\partial_\nu {\mathcal A}_\mu\right), \nonumber \\ 
v^{(s)}_{\mu\nu}&=&\partial_\mu v^{(s)}_\nu-\partial_\nu v^{(s)}_\mu=-e\frac{I}{2}\left(\partial_\mu {\mathcal A}_\nu-\partial_\nu {\mathcal A}_\mu\right)\,. 
\label{eq:fvmunu}
\eea 
Since the bremsstrahlung cross section is proportional to $\alpha/m^2$ for a radiating particle of mass $m$, 
we expect the NLO contributions to be small compared to the LO due to the large proton mass.    
%
\begin{figure}[b]
\centering
         \includegraphics[width=\columnwidth,height=3cm]{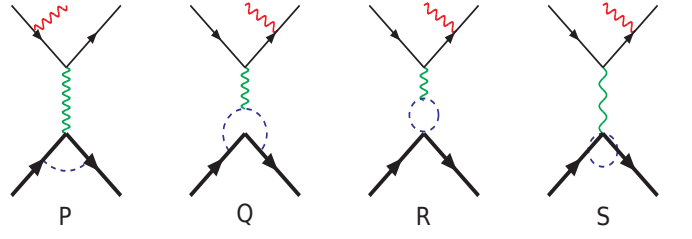}
\caption{A subset of all NNLO Feynman diagrams of the lepton-proton bremsstrahlung process, where the 
pion loops, shown by the dashed (blue) propagator lines, contribute to the proton form factors. 
The vertices in the NNLO pion loop diagrams are all generated by ${\cal L}_{\pi N}^{(0)}$. 
}
\label{fig:NNLOfeyndiag}
\end{figure}
%

In $\chi$PT we naively expect that the NNLO corrections are of order $(\mathscr{P}/\Lambda_\chi)^3\ll1$,
where $\mathscr{P}$ denotes the four-momentum of the process of the order of the pion mass, $m_\pi$. However, 
it turns out that the pion loop contributions are as important as the lower chiral order nucleon pole graph 
contributions. We will investigate the question regarding the magnitudes of the NNLO pion loop contributions 
in our process, i.e., we examine how well the usual chiral counting works for the lepton-proton bremsstrahlung 
process in Sec.~\ref{section:summary}. At NNLO the Lagrangian  ${\cal L}_{\pi N}^{(2)}$, can be written as 
${\cal L}_{\pi N}^{(2)}={\mathscr{L}}^{(2),{\rm fixed}}_{\pi N}+{\mathscr{L}}^{(2)}_{\pi N}$, 
where ${\mathscr{L}}^{(2),{\rm fixed}}_{\pi N}$ 
contains $m_p^{-2}$ recoil terms with known coefficients, while ${\mathscr{L}}^{(2)}_{\pi N}$ 
contains low-energy constants (LECs). 
The LECs regularize the ultra-violet (UV) contributions from the pion loop diagrams, which enter at NNLO order. The 
pion loops associated with the photon-proton vertex  in diagrams (A) and (B) of Fig.~\ref{fig:LOfeyndiag} are 
illustrated in Fig.~\ref{fig:NNLOfeyndiag}. They contribute to the first momentum dependence of the 
proton form factors in $\chi$PT. 
In this work we will effectively make use of the NNLO proton electric form factor which along with the magnetic form 
factor was formally derived in Refs.~\cite{fettes00,Bernard1998,hemmert1998,Ecker1994} using $\chi$PT. The explicit 
${\cal L}_{\pi N}^{(2)}$ expression can also be found in those references. At this chiral order the measured r.m.s. charge 
and magnetic radii determine LECs in ${\cal L}_{\pi N}^{(2)}$~\cite{Bernard1998}. In essence, a measure of the NNLO 
contributions to our cross section is the following electric Sachs form factor $G^p_E$, expressed in terms of the 
corresponding iso-vector $G^v_E$ and iso-scalar $G^s_E$ form factors, namely 
\bea
G^p_{E}(q^2)=\frac{1}{2}\left[G^v_E(q^2) + G^s_E(q^2)\right] \approx 1 + \frac{q^2}{6}(r^p_{E})^2 + {\mathcal O}(q^4),
\nonumber\\
\label{eq:GEM} 
\eea
where $r^p_{E}$ is the electric radius of the proton. 
In Sec.~\ref{section:summary}, our rough estimate of the NNLO contribution is obtained by folding the LO 
differential cross section results with $G^p_E$ which include the ``measured'' r.m.s. radius as the 
phenomenological input. These should {\it effectively simulate} the NNLO $\chi$PT pion loop contributions.

\section{The LO and NLO cross sections}
\label{section:cross-sections}

At LO only the Feynman diagrams (A) and (B) in Fig.~\ref{fig:LOfeyndiag}, contribute. We denote the incident and 
scattered lepton four-momenta as $p_l=(E_l,\vec{p})$ and $p^\prime_l=(E^\prime_l,\vec{p}^{\,\prime})$, respectively, where, 
e.g.,  $E_l=\sqrt{m_l^2+\vec{p}^{\, 2}}$. The corresponding proton four-momenta are $P_p=(E_p,\vec{P}_p)$ and 
$P^\prime_p=(E^\prime_p,\vec{P}^{\,\prime}_p)$, and the outgoing photon has the four-momentum $k=(E_\gamma,\vec{k})$. 
Furthermore, $\theta$ is the lepton scattering angle such that  
$\vec{p}\cdot\vec{p}^{\, \prime} =  |\vec{p}\, | |\vec{p}^{\, \prime}| {\rm cos} \theta$, and  $q=(Q - k)$ is the 
four-momentum transferred to the proton when the lepton is radiating.  

\begin{figure}[t]
\centering
         \includegraphics[scale=0.55]{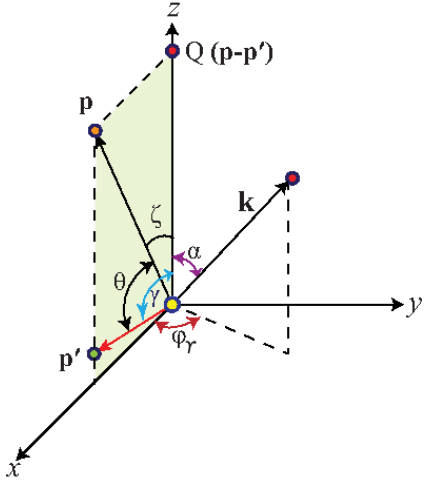}
\caption{Reference coordinate system used in the evaluation of the differential cross section, 
such that $\vec{Q} = \vec{p}-\vec{p}^{\,\prime}$ is taken along the $z$-axis, while $\vec{p}$ and 
$\vec{p}^{\, \prime}$ lie in $xz$-plane. The different angles and three-momentum symbols are defined 
in the text.}
\label{fig:kin_ref}
\end{figure}

In $\chi$PT the non-relativistic heavy proton four-momentum satisfies re-parametrization 
invariance~\cite{Bernard1992,Manohar}, and takes the form $P^\mu_p = m_p v^\mu+p_p^\mu$, where $m_p$ is the proton mass, 
such that the square of its off-shell {\it residual} part is $p_p^2\ll m_p^2$.  This means that the incident proton 
kinetic energy to lowest order in $m_p^{-1}$ becomes  $v\cdot p_p = \frac{\vec{p}_p^{\, 2}}{2m_p} +\cdots$, and similarly 
for the final recoiling proton. The bremsstrahlung differential cross section is given in the laboratory frame by the 
general expression 
\bea
     {\rm d}\sigma  
      &=&\int\frac{ {\rm d}^3 \vec{p}^{\,\prime} \; {\rm d}^3 \vec{k} }{ (2\pi)^5 8 E_l^\prime E_\gamma} 
      \frac{  \delta \left(E_l-E_l^\prime-E_\gamma   - 
      \frac{ ( \vec{p}-\vec{p}^{\, \prime}-\vec{k})^2}
      {2m_p}+\cdots\right) }{4m_p E_l \left(m_p + \frac{ (\vec{p}
      -\vec{p}^{\, \prime}-\vec{k})^2}{2m_p}+\cdots\right) } \, 
\nonumber\\ && \nonumber\\
      &&\quad \times\,\, \frac{1}{4}\sum_{\rm spins} | {\mathcal M}_{br}|^2 \, , 
\label{eq:sigma1}
\eea
where in the phase-space expression (including the $\delta$-function) we expand the recoil proton energy 
as\footnote{For a $\chi$PT analysis for this process up to and including NNLO, it is sufficient to expand 
kinematic quantities up to ${\mathcal O}(m^{-2}_p)$.} 
\bea
      E^\prime_p&=&\sqrt{m^2_p+(\vec{p}^{\,\prime}_p)^2} \nonumber \\
              &=&m_p+\frac{( \vec{p}-\vec{p}^{\,\prime}-\vec{k})^2}{2m_p} +{\mathcal O}\left(m^{-3}_p\right) \, .  
\label{eq:Eprimerecoil}
\eea 
A straightforward evaluation of the two Feynman diagrams (A) and (B), shown in Fig.~\ref{fig:LOfeyndiag} 
leads to the following LO expression for the bremsstrahlung amplitude squared:  
\begin{widetext}
\bea
     \frac{1}{4}\sum_{\rm spins} \!\! |{\mathcal M}^{\rm (LO)}_{br}|^2 &=& \frac{1}{4}\sum|{\mathcal M}_{A}+{\mathcal M}_{B}|^2 = \left(\frac{512\pi^3\alpha^3}{q^4}\right)(m_p+E_p)(m_p+E_p^\prime) 
  \nonumber \\ 
     &&\times\,\Biggl\{ -\,\, \frac{1}{[(p^\prime+k)^2-m_l^2]^2} 
     \left[ m_l^4 + m_l^2E_l^\prime E_l  + m_l^2 E_l E_\gamma 
     + m_l^2 E_l^\prime E_\gamma - E_l E_l^\prime E_\gamma^2 + m_l^2 (\vec{k}\cdot\vec{p}) 
     - m_l^2 (\vec{k}\cdot\vec{p}^{\, \prime}) 
     \right. 
  \nonumber \\  
     && \left. \hspace{1.74in}
     + \,\, m_l^2 (\vec{p}\cdot\vec{p}^{\, \prime}) + E_l E_\gamma (\vec{k}\cdot\vec{p}^{\, \prime}) 
     - E_l^\prime E_\gamma  (\vec{k}\cdot\vec{p}) + (\vec{k}\cdot\vec{p}^{\, \prime})(\vec{k}\cdot\vec{p}) 
     \right]
  \nonumber \\ 
     && \hspace{0.55cm} - \,\, \frac{1}{[(p-k)^2-m_l^2]^2} 
     \left[ m_l^4 - E_\gamma^2 E_l E_l^\prime + m_l^2 E_l E_l^\prime 
     - m_l^2 E_l E_\gamma - m_l^2E_l^\prime E_\gamma + m_l^2 (\vec{p}\cdot\vec{k}) + m_l^2 (\vec{p}\cdot\vec{p}^{\, \prime}) 
     \right. 
  \nonumber \\ 
     && \left. \hspace{1.74in}
     - \,\, m_l^2 (\vec{k}\cdot\vec{p}^{\, \prime}) - E_l E_\gamma (\vec{k}\cdot\vec{p}^{\, \prime}) 
     + E_l^\prime E_\gamma (\vec{k}\cdot\vec{p}) + (\vec{k}\cdot\vec{p}^{\, \prime})(\vec{k}\cdot\vec{p})  
     \right] 
  \nonumber \\ 
     && \hspace{0.55cm} - \,\, \frac{2}{[(p^\prime+k)^2-m_l^2]\,[(p-k)^2-m_l^2]}
     \left[  m_l^2 E_l^\prime E_l -  m_l^2E_\gamma^2 + E_l^2 E_l^{\prime 2} - m_l^2 (\vec{p}\cdot\vec{p}^{\, \prime}) 
     + E_l^2 (\vec{k}\cdot\vec{p}^{\,\prime}) 
     \right. 
  \nonumber \\ 
     && \left. \hspace{1.74in}
     - \,\, E_l^{\prime \, 2} (\vec{k}\cdot\vec{p}) - (\vec{p}\cdot\vec{p}^{\, \prime})^2 
     + (\vec{p}\cdot\vec{p}^{\,\prime})(\vec{k}\cdot\vec{p}^{\,\prime}) - (\vec{p}\cdot\vec{p}^{\,\prime})(\vec{k}\cdot\vec{p}\,) 
     \right] \,\, \Biggr\}\,. 
\label{eq:sigma3}
\eea  
\end{widetext}
To evaluate the cross section, it is convenient to define our reference frame such that the momentum transfer, 
$\vec{Q}=\vec{p}-\vec{p}^{\, \prime}$ is directed along the $z$-axis~\cite{motsai1969}, while the lepton momenta, 
$\vec{p}$ and $\vec{p}^{\, \prime}$, lie in $xz$-plane as shown in Fig.~\ref{fig:kin_ref}. The pertinent angles 
are defined as follows:  
\bea
     \vec{k}\cdot\vec{p}^{\,\prime} &=& E_\gamma|\vec{p}^{\,\prime}|(\cos\gamma\,\cos\alpha
     +\sin\alpha\,\sin\gamma\,\cos\phi_\gamma), 
\nonumber\\ 
     \vec{k}\cdot\vec{p} &=& E_\gamma|\vec{p}\,|(\cos\zeta\,\cos\alpha
     +\sin\alpha\,\sin\zeta\,\cos\phi_\gamma)\, . 
\eea 
The lepton scattering angle in our coordinate system is given as $\theta =\gamma - \zeta$. 
When the lepton radiates a photon the squared four-momentum transferred to the proton is 
\bea
    q^2 &=& (Q-k)^2 \nonumber \\
        &=& 2\biggl[m^2_l - E_lE_l^\prime+|\vec{p}\,||\vec{p}^{\, \prime}|\cos\theta - E_\gamma(E_l-E_l^\prime) \nonumber \\ 
        && +\, E_\gamma\cos\alpha \sqrt{|\vec{p}\,|^2+|\vec{p}^{\, \prime}|^2-2|\vec{p}\,||\vec{p}^{\, \prime}|\cos\theta}\biggr]\, ,  
\label{eq:q_squared}
\eea
which is independent of $\phi_\gamma$ in our reference frame as suggested in Ref.~\cite{motsai1969}.   
This choice of reference frame readily allows the analytical $\phi_\gamma$ integration.  
%
%

To evaluate our LO expression for the  differential  cross section, 
${\rm d}^3\sigma^{\rm(LO)}/({\rm d} |\vec{p}^{\, \prime}|\,{\rm d}\Omega^\prime_l\,{\rm d}{\rm cos}\alpha)$ 
[see Eq.~(\ref{eq:sigma5})], it is convenient to define two angle dependent parameters,
\begin{equation*}
a= \frac{1-\beta^\prime  \cos\alpha\, \cos\gamma }{\beta^\prime \sin\alpha\, \sin\gamma}\quad 
{\rm and}\quad 
b= \frac{1-\beta \cos\alpha\, \cos\zeta }{\beta \sin\alpha\, \sin\zeta}\,.
\end{equation*}
Here $\beta=\frac{|\vec{p} \, | }{E_l}$ and $\beta^\prime=\frac{|\vec{p}^{\, \prime} |}{E_l^\prime}$ 
are the incoming and outgoing lepton velocities, respectively. In this equation, we also define the magnitudes 
of the incoming and outgoing lepton three-momenta as $p=|\vec{p}|$ and $p^\prime = |\vec{p}^{\,\prime}|$, respectively. 
To obtain our final expression, we integrate the cross section in Eq.~(\ref{eq:sigma1}) over the photon energies  
$E_\gamma$, which means that the infrared singularity will appear when the momentum $p^\prime$ is close to it's maximal 
allowed value. To account for the proton recoil corrections in the kinematics we include the $m_p^{-1}$ terms in the 
photon energy $E_\gamma$ given by the delta-function in Eq.~(\ref{eq:sigma1}). We define $E_\gamma^0=E_l-E_l^\prime $, 
which gives the following expression for $E_\gamma$, 
\begin{eqnarray} 
E_\gamma = E_\gamma^0 -\frac{K}{m_p}+{\mathcal O}\left(m^{-3}_p\right) \, ,  
\label{eq:E_gamma}
\end{eqnarray}
where
\bea
K&=& \frac{1}{2}\biggl[|\vec{p}|^2+|\vec{p}^{\,\prime}|^2-2|\vec{p}||\vec{p}^{\,\prime}|\cos\theta+(E_\gamma^0)^2 \nonumber \\
&& -\, 2 E_\gamma^0 \cos\alpha \sqrt{|\vec{p}|^2+|\vec{p}^{\,\prime}|^2-2|\vec{p}||\vec{p}^{\,\prime}|\cos\theta}\biggr]\,.
\eea
When integrating over $E_\gamma$, the expression in the delta-function also introduces a factor 
$\left(1-\frac{Z}{m_p}\right)$ in the phase-space to lowest order in $m_p^{-1}$ needed in our analysis, where 
\beq
Z= E_\gamma^0 -  \cos\alpha \, \sqrt{|\vec{p}|^2+|\vec{p}^{\,\prime}|^2-2|\vec{p}||\vec{p}^{\,\prime}|\cos\theta}. 
\eeq
Furthermore, the $m_p^{-1}$ correction in the photon energy affects the expression for the four-momentum transfer 
$q^2$, which can be written as 
\beq
q^2= (q^0)^2+\frac{\kappa}{m_p}+{\mathcal O}\left(m^{-3}_p\right)\,,
\eeq 
where 
\bea
(q^0)^2 &=& 2\biggl[ m^2_l - E_lE_l^\prime+|\vec{p}\,||\vec{p}^{\, \prime}|\cos\theta - E_\gamma^0(E_l-E_l^\prime) \nonumber \\
&& +\, E_\gamma^0\cos\alpha \sqrt{|\vec{p}\,|^2+|\vec{p}^{\, \prime}|^2-2|\vec{p}\,||\vec{p}^{\, \prime}|\cos\theta}\biggr], 
\label{eq:q0_squared}
\eea
and 
\bea 
\kappa = 2 K\Big[ E_l-E_l^\prime -\cos\alpha \sqrt{|\vec{p}\,|^2
+|\vec{p}^{\, \prime}|^2-2|\vec{p}\,||\vec{p}^{\, \prime}|\cos\theta}\Big]\,.\nonumber \\
\eea 
In the process of evaluation of the cross section up to NNLO, we will only need  the $m_p^{-1}$ kinematic terms, 
i.e.,  we include the $m_p^{-1}$ corrections for $E_\gamma$, $q^2$, as well as in the above given phase-space factor.  

As a pedagogical survey of the bremsstrahlung process at energies not much larger than the muon mass, 
we initially consider the most simplest and rather qualitative case of the static proton limit ($m_p\to \infty$). 
We shall then compare these qualitative results with the improved ones obtained by, first, including the kinematical 
$m_p^{-1}$ recoil corrections in the phase space factor and the delta-function expression in Eq.~(\ref{eq:sigma1}), 
and second, by including the dynamical $m_p^{-1}$ recoil corrections in the matrix element at NLO. 

As a first step, we evaluate the LO cross section in the static proton limit ($m_p\to\infty$), which is expressed as
\bea
     {\rm d}\sigma^{\rm(LO)}_{{\rm static}} &=& \int  
     \frac{ {\rm d}^3\vec{p}^{\, \prime}\,  {\rm d}^3 \vec{k} }{ (2\pi)^5 32 E_l^\prime E_\gamma m_p^2 E_l } 
     \delta \left(E_l-E_l^\prime-E_\gamma  \right) 
\nonumber\\ 
     && \quad \times\,\, \frac{1}{4}\sum_{spin} | {\mathcal M}_{br}^{\rm static}|^2 \,.   
\label{eq:sigma2}
\eea
When we evaluate ${\mathcal M}_{br}^{\rm static}$ in Eq.~(\ref{eq:sigma2}) in the static proton limit ($m_p\to\infty$), 
we  set $K$, $\kappa$ and $Z$ all equal zero in Eq.~(\ref{eq:sigma5}). This equation incorporates the $m_p^{-1}$ recoil 
effects only from the phase space [including the energy-delta function in Eq.~(\ref{eq:sigma1})] but the matrix element 
is derived from the leading chiral order Lagrangian ${\mathcal L}^{(0)}_{\pi N}$:
\begin{widetext}
\bea
     \frac{{\rm d}^3\sigma^{\rm(LO)}}{  {\rm d}p^{\prime}\, {\rm d}\Omega^\prime_l\, {\rm d} \cos\alpha } &=&
     \left(\frac{\alpha^3  \beta^{\prime\, 2} }{2\pi^2q^4}\right)\left( 1-\frac{Z}{m_p}\right) E_\gamma 
  \nonumber \\ 
     &&\left\{
     - \int_0^{2\pi} {\rm d}\phi_\gamma \frac{1}{(a- \cos\phi_\gamma)^2} 
     \left(\frac{1}{E_lE^\prime_lE_\gamma^2(\beta^{\prime}\, {\sin}\alpha\, {\sin}\gamma)^2}\right)\right.
  \nonumber \\ 
     &&\left. \hspace{0.8cm}
     \times\, \Big[ m_l^4 - E_l E_l^\prime E_\gamma^2 + m_l^2 E_l^\prime E_\gamma  
     + m_l^2 E_l E_l^\prime  + m_l^2 E_l E_\gamma + m_l^2 p p^\prime \cos\theta \right. 
  \nonumber \\ 
     &&\left. \hspace{1.3cm}
     -\, m_l^2 p^\prime E_\gamma\, \Big({\cos}\alpha\, {\cos}\gamma 
     +\, {\sin}\alpha\, {\sin}\gamma\, {\cos}\phi_\gamma \Big)
     +\, p^\prime E_l E_\gamma^2\, \Big({\cos}\alpha\, {\cos}\gamma 
     +\, {\sin}\alpha\, {\sin}\gamma\, {\cos}\phi_\gamma \Big) \right.
  \nonumber \\ 
     &&\left. \hspace{1.3cm}
     +\, m_l^2 p E_\gamma\, \Big( {\cos}\alpha\, {\cos}\zeta
     +\, {\sin}\alpha\, {\sin}\zeta\, {\cos}\phi_\gamma \Big)
     -\, p E_l^\prime E_\gamma^2\, \Big({\cos}\alpha\, {\cos}\zeta
     +\, {\sin}\alpha\, {\sin}\zeta\, {\cos}\phi_\gamma \Big) \right.
  \nonumber \\ 
     &&\left. \hspace{1.3cm}
     +\, p p^\prime E_\gamma^2\, \Big( {\cos}^2\alpha\, {\cos}\gamma\, {\cos}\zeta 
     +\, {\cos}\alpha\, {\cos}\gamma\, {\sin}\alpha\, {\sin}\zeta\, {\cos}\phi_\gamma
     +\, {\sin}\alpha\, {\sin}\gamma\, {\cos}\alpha\, {\cos}\zeta\, {\cos}\phi_\gamma \right.
  \nonumber \\ 
     &&\left. \hspace{1.3cm}
     +\, {\sin}^2\alpha\, {\sin}\gamma\, {\sin}\zeta\, {\cos}^2\phi_\gamma \Big)\Big]_{{\rm dir}(\gamma)} \right.
  \nonumber \\ 
     &&\left.
     -\, \int_0^{2\pi} {\rm d}\phi_\gamma \frac{1}{(b- \cos\phi_\gamma)^2}
     \left( \frac{ E_l^\prime }{E_l^3 E_\gamma^2 (\beta\, {\sin}\alpha\, {\sin}\zeta)^2}\right) \right.
  \nonumber \\
     &&\left. \hspace{0.8cm}
     \times\, \Big[ m_l^4 - E_l E_l^\prime E_\gamma^2 - m_l^2 E_l^\prime E_\gamma + m_l^2 E_l E_l^\prime - m_l^2 E_l E_\gamma 
     + m_l^2 p p^\prime \cos\theta \right.
  \nonumber \\ 
     &&\left. \hspace{1.3cm}
     -\, m_l^2 p^\prime E_\gamma\,\Big( {\cos}\alpha\, {\cos}\gamma 
     +\, {\sin}\alpha\, {\sin}\gamma\, {\cos}\phi_\gamma\Big) 
     -\, p^\prime E_l  E_\gamma^2\,  \Big( {\cos}\alpha\, {\cos}\gamma 
     +\, {\sin}\alpha\, {\sin}\gamma\, {\cos}\phi_\gamma \Big) \right.
  \nonumber \\ 
     &&\left.  \hspace{1.3cm}    
     +\, m_l^2 p E_\gamma\,\Big( {\cos}\alpha\, {\cos}\zeta 
     +\, {\sin}\alpha\, {\sin}\zeta\, {\cos}\phi_\gamma \Big)
     +\, p E_l^\prime  E_\gamma^2\,  \Big( {\cos}\alpha\, {\cos}\zeta 
     +\, {\sin}\alpha\, {\sin}\zeta\, {\cos}\phi_\gamma\Big) \right.
  \nonumber \\ 
     &&\left. \hspace{1.3cm}
     +\, p p^\prime E_\gamma^2\, \Big({\cos}^2\alpha\, {\cos}\gamma \, {\cos}\zeta 
     +\, {\cos}\alpha\, {\cos}\gamma\, {\sin}\alpha\, {\sin}\zeta\, {\cos}\phi_\gamma 
     +\, {\sin}\alpha\, {\sin}\gamma \,{\cos}\alpha\, {\cos}\zeta\, {\cos}\phi_\gamma\right.
  \nonumber \\ 
     &&\left. \hspace{1.3cm}
     +\,  {\sin}^2\alpha\, {\sin}\gamma \, {\sin}\zeta\, {\cos}^2\phi_\gamma \Big)
     \Big]_{{\rm dir}(\zeta)} \right.
  \nonumber \\ 
     &&\left. 
     +\, \int_0^{2\pi} {\rm d}\phi_\gamma \frac{1}{(a- \cos\phi_\gamma)(b- \cos\phi_\gamma) }
     \left( \frac{ 2 }{E_l^2 E_\gamma^2 (\beta \beta^\prime\, {\sin}^2\alpha\, {\sin}\gamma\, {\sin}\zeta) }\right) \right.
  \nonumber \\ 
     &&\left. \hspace{0.8cm}
     \times\, \Big[ m_l^2 E_l E_l^\prime + E_l^2 E_l^{\prime 2} - m_l^2 E_\gamma^2 
     -\, m_l^2 p p^\prime \cos\theta  - p^2 p^{\prime 2}\, {\cos}^2\theta \right.
  \nonumber \\ 
     &&\left. \hspace{1.3cm}
     +\, p^\prime E_l^2 E_\gamma\, \Big( {\cos}\alpha\, {\cos}\gamma
     +\, {\sin}\alpha\, {\sin}\gamma\, {\cos}\phi_\gamma \Big) 
     +\, p p^{\prime 2} E_\gamma  \cos\theta\,   \Big({\cos}\alpha\, {\cos}\gamma 
     +\, {\sin}\alpha\, {\sin}\gamma\, {\cos}\phi_\gamma \Big)\right.
  \nonumber \\ 
     &&\left. \hspace{1.3cm}
     -\, p E_l^{\prime 2} E_\gamma\,  \Big( {\cos}\alpha\, {\cos}\zeta 
     +\, {\sin}\alpha\, {\sin}\zeta\, {\cos}\phi_\gamma \Big) 
     -\, p^2 p^{\prime} E_\gamma  \cos\theta\, \Big( {\cos}\alpha\, {\cos}\zeta 
     +\, {\sin}\alpha\, {\sin}\zeta\, {\cos}\phi_\gamma \Big) 
     \Big]_{\rm int} \textcolor{white}{\int_0^{2\pi}\hspace{-0.6cm}}\right\}\, .\nonumber \\
\label{eq:sigma5}
\eea 
\end{widetext}
The terms within the first and second square brackets, i.e., $[\cdots]_{{\rm dir}(\gamma)}$ and 
$[\cdots]_{{\rm dir}(\zeta)}$, represent the contributions from the ``direct'' terms (matrix element 
squared  of diagram (B) and (A), respectively) of real photon emissions from the outgoing and 
incoming leptons, respectively. The third square bracket $[\cdots]_{\rm int}$, represents  the   
``interference'' contribution of diagrams (A) and (B).  

Next including the ${\mathcal O}(m^{-1}_p)$ dynamical corrections in the matrix elements due to the 
interactions in ${\cal L}^{(1)}_{\pi N}$, i.e., with 
\bea
\left|\mathcal{M}_{br}\right|^2 &\to& 2 {\rm Re}\sum_{\rm spins}\left({\mathcal M}_{\rm A}+{\mathcal M}_{\rm B}\right)^{*} 
\nonumber \\
&& \qquad \times\,\left({\mathcal M}_{\rm E}+{\mathcal M}_{\rm F}+{\mathcal M}_{\rm G}+{\mathcal M}_{\rm H}\right)\,, 
\label{eq:NLO_amps}
\eea
yields the complete NLO expression to the bremsstrahlung differential cross section which is expressed as 
\bea
\frac{{\rm d}^3\sigma}{  {\rm d}p^{\,\prime}\, {\rm d}\Omega^\prime_l\, {\rm d} \cos\alpha } &=&\frac{{\rm d}^3\sigma^{\rm(LO)}}{  {\rm d}p^{\,\prime}\, {\rm d}\Omega^\prime_l\, {\rm d} \cos\alpha } + \Delta\left[\frac{\rm d^3 \sigma}{{\rm d} p^{\prime}\, {\rm d} \Omega_l^{\prime}\, {\rm d} \cos\alpha}\right]_{\rm NLO},\nonumber \\
\label{eq:sigma6}
\eea
where the $\mathcal O(m^{-1}_p)$ NLO correction term above is   
\begin{widetext}
\bea
\Delta\left[\frac{\rm d^3 \sigma}{{\rm d} p^{\prime}\, {\rm d} \Omega_l^{\prime}\, {\rm d} \cos\alpha}\right]_{\rm NLO}\!\!\!&=&\left(\frac{\alpha^3 \beta^{\prime 2}}{2\pi^2(q^0)^4}\right) \frac{1}{m_p}2E^0_\gamma
\nonumber \\
&&\times\, \int^{2\pi}_0 {\rm d} \phi_\gamma 
\Bigg[-\frac{1}{(a-\cos\phi_\gamma)^2}\left(\frac{W_{\rm BF}}{E_lE_l^{\prime}(E^0_\gamma)^2(\beta^{\prime}\, \sin\alpha\, \sin\gamma)^2}\right) 
\nonumber\\
&&\hspace{1.95cm}-\, \frac{1}{(b-\cos\phi_\gamma)^2}\left(\frac{E^{\prime}_lW_{\rm AE}}{E_l^3(E^0_\gamma)^2(\beta\, \sin\alpha\, \sin\zeta)^2}\right)
\nonumber\\
&&\hspace{1.95cm}-\,\, \frac{1}{(a-\cos\phi_\gamma)(b-\cos\phi_\gamma)}\left(\frac{W_{\rm AF}+W_{\rm BE}}{E^2_l (E^0_\gamma)^2(\beta \beta^\prime\, {\sin}^2\alpha\, \sin\gamma\, \sin\zeta)}\right)
\nonumber\\
&&\hspace{1.95cm}+\,\, \frac{(q^0)^2}{Q^2}\bigg\{\frac{1}{(a-\cos\phi_\gamma)}\left(\frac{W_{\rm BG}-W_{\rm BH}}{E_lE_l^{\prime} (E^0_\gamma)^2(\beta^{\prime}\, \sin\alpha\, \sin\gamma)}\right)
\nonumber \\
&&\hspace{3.4cm}-\,\,\frac{1}{(b-\cos\phi_\gamma)}\left(\frac{E^{\prime}_l(W_{\rm AG}+W_{\rm AH})}{E^3_l (E^0_\gamma)^2(\beta\, \sin\alpha\, \sin\zeta)}\right)\bigg\}\,\,\Bigg]\,.\nonumber\\
\label{eq:sigma7}
\eea
\end{widetext} 
At NLO we use the $\mathcal O(m^{0}_p)$ expressions, $E^0_\gamma = E_l - E_l^\prime$ and $(q^0)^2$, 
already defined earlier in Eq.~\eqref{eq:E_gamma} and Eq.~\eqref{eq:q0_squared}, respectively. 
The explicit expressions for the partial amplitudes of $\mathcal O(m^{0}_p)$, namely 
$W_{\rm AE},W_{\rm AF},W_{\rm AG},W_{\rm AH}$ and $W_{\rm BE},W_{\rm BF},W_{\rm BG},W_{\rm BH}$, are rather lengthy 
and relegated to the Appendix. As evident from the subscripts, these contributions arise from the 
interference of the diagrams in Fig.~\ref{fig:LOfeyndiag} with those in Fig.~\ref{fig:NLOfeyndiag}. 

\section{LO and NLO results} 
\label{section:results_LO+NLO} 
The MUSE collaboration~\cite{gilman13} proposes the  scattering of lepton off proton at the following three 
beam momenta, $p$ = 115, 153 and 210 MeV/c. As discussed, MUSE is designed to count the number of scattered 
leptons at a fixed scattering angle $\theta$ for any value of the scattered lepton momentum $|\vec{p}^{\,\prime}|$ 
larger than a certain minimum value.  We shall discuss the dependence of the cross section on the photon angle 
$\alpha$ and the lepton momentum $|\vec{p}^{\,\prime}|$. First, however, we analyze the $q^2$ dependence of the 
bremsstrahlung process. 

In order to extract a precise value for the proton r.m.s. radius, one needs to know accurately the $Q^2$ 
dependence of the proton form factor. To LO in $\chi$PT the four-momentum transferred to the proton is 
$q=(Q-k)$, since the proton do not radiate. At NLO, however, the momentum transfer can be 
either $Q$ or $q$ depending on whether the proton radiates or not. For a given scattering angle $\theta$, 
Eq.~(\ref{eq:q_squared}) shows that  $q^2$ is  a function of the outgoing lepton momentum $|\vec{p}^{\,\prime}|$, 
the lepton scattering angle $\theta$, and the photon polar angle $\alpha$. Although the bremsstrahlung process for 
muon scattering at a given angle $\theta$, constitutes a small correction to the elastic cross section, the process 
introduces a non-negligible $q^2$ value uncertainty. Thus, we find it important to examine the $q^2$ dependence 
on $|\vec{p}^{\,\prime}|$ and $\alpha$ in order to guesstimate the uncertainty given by the bremsstrahlung process. 

\begin{figure}[t]
\centering
         \includegraphics[scale=0.3]{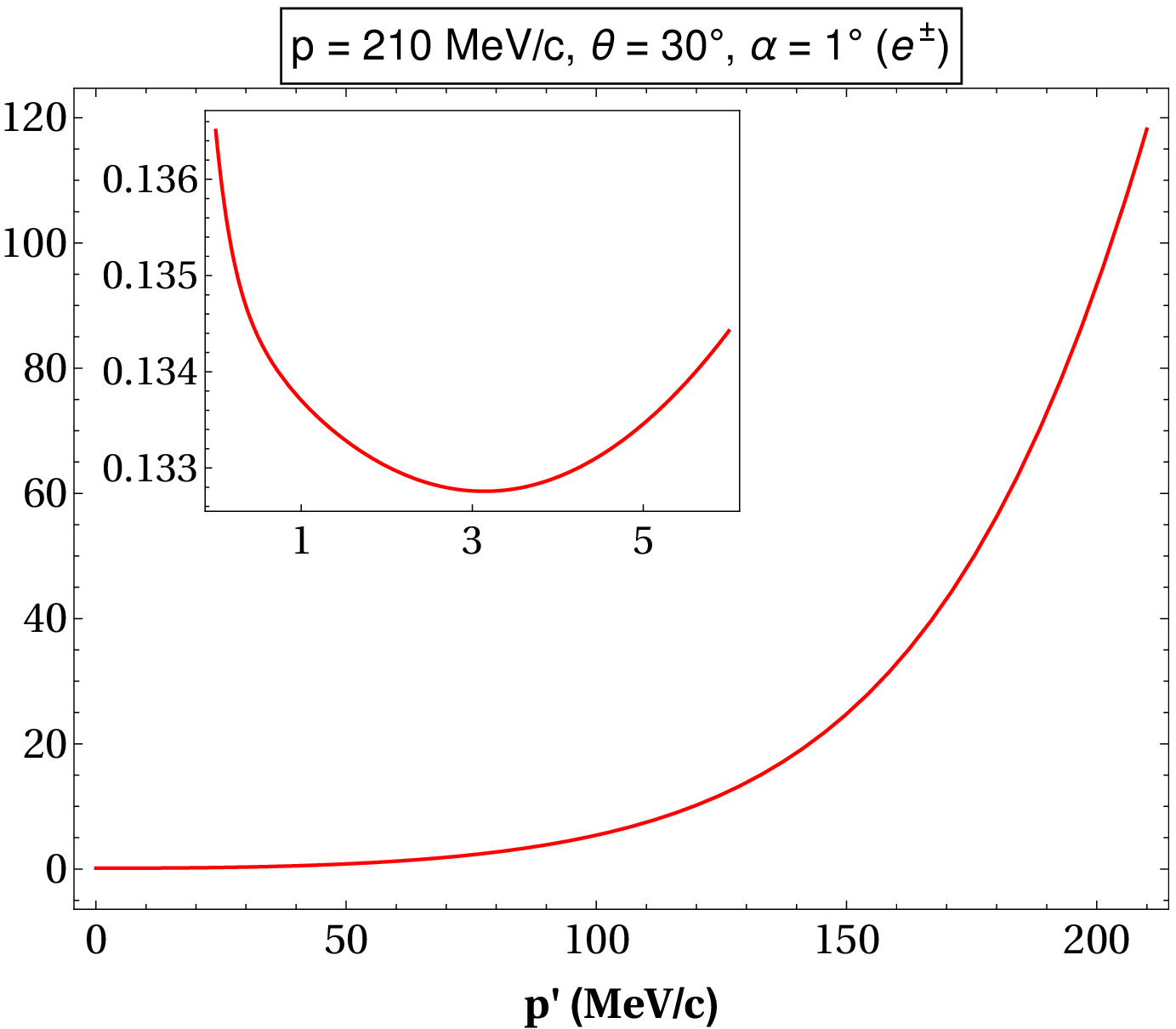}  \includegraphics[scale=0.3]{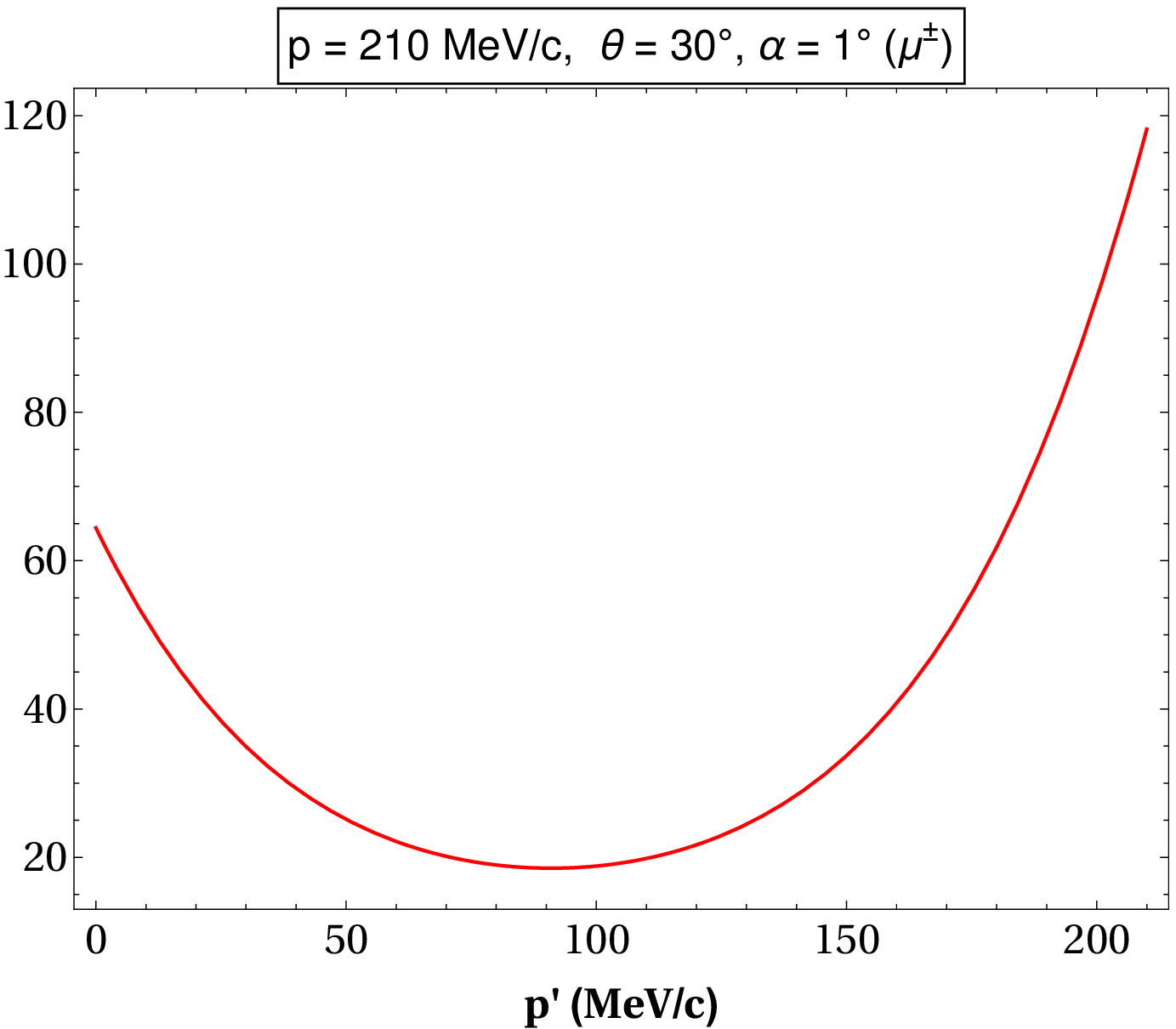}
\caption{Behavior of the squared momentum transfer $-q^2$ (in units of $10^2$ MeV${}^2/c^2$), as a 
function of the outgoing lepton momentum, $p^{\prime}=|\vec{p}^{\,\prime}|$, for a fixed incident lepton 
momenta, $p=|\vec{p}\,|=210$ MeV/c, scattering angle, $\theta=30\degree$ and $\alpha=1\degree$. When 
$p^\prime \ll m_e , |\vec{p}\, |$, we have near collinear photon emission with $\vec{p}$. In the case of 
electron scattering (left plot), the inset plot clearly displays the minimum of $-q^2$. }
\label{fig:momentum_transfer}
\end{figure}

First, it is inferred from  Eq.~\eqref{eq:q_squared} that for a given lepton mass $m_l$ and very small 
$|\vec{p}^{\,\,\prime}| = p^\prime$, as the angle $\alpha\to 0$, the $\theta$ angular dependence of the squared momentum 
transfer $-q^2$ becomes practically negligible. This means that the photon emission direction $\vec{k}$ is (almost) 
{\it collinear} with the incident lepton direction $\vec{p}$. 
On the other hand, when $p^{\prime}$ tends toward its maximal limit for fixed $\theta$, i.e., 
$p^{\prime}\to p^{\prime}_{elastic}(\theta)$, the $\alpha$ dependence plays a complex role in determining the resulting 
$-q^2$ behavior. 
Figure~\ref{fig:momentum_transfer} depicts the $-q^2$ behavior of the outgoing lepton momentum $p^{\prime}$, for a fixed 
incident lepton momentum $p=|\vec{p}\, |$, forward scattering angle $\theta$, and a small polar angle $\alpha$. Both 
plots exhibit a quadratic behavior of $-q^2$ versus $p^{\prime}$, with a minimum at a certain $p^{\prime}$ value. In general, 
the minimum depends on $p$, $\theta$, $\alpha$ and the lepton mass $m_l$. Even in the massless ($m_l\to 0$) case, a minimum 
of $-q^2$ at a non-zero value of $p'$ is obtained as long as $\theta$ or $\alpha$ is non-zero. For example, in the given 
figure the minimum occurs for $p^\prime\lesssim 5$~MeV/c for the electron and $p^\prime\lesssim 100$~MeV/c for the muon. 
Furthermore, we find that for given fixed angles ($\theta,\alpha$), and lepton mass $p^{\prime}\ll m_l \leq p^{\prime}_{elastic}$, 
the square momentum transfer $-q^2$ becomes linear in $|\vec{p}^{\,\prime}|$ with a negative slope for forward scattering angles 
$\theta<\pi/2$. This behavior can be seen in each plot in Fig.~\ref{fig:momentum_transfer} for the small $p^\prime$ region, 
though in case of the electron plot the negative slope of the hardly discernible. However, the inserted zoomed plot clearly 
shows this behavior. Thus, we can expect that the small $q^2$ dependence on $p^{\prime}$ in the low-momentum region below 100 
MeV/c, that is relevant to MUSE, will produce significant effects on the differential cross section 
d$\sigma$/(d$p^\prime$d$\Omega^\prime)$. Note that the MUSE collaboration is expected to detect electrons and muons with momenta 
$p^{\prime}$ in a range down to about $50-20$ MeV/c.

\begin{figure*}[t]
\centering
         \includegraphics[scale=0.36]{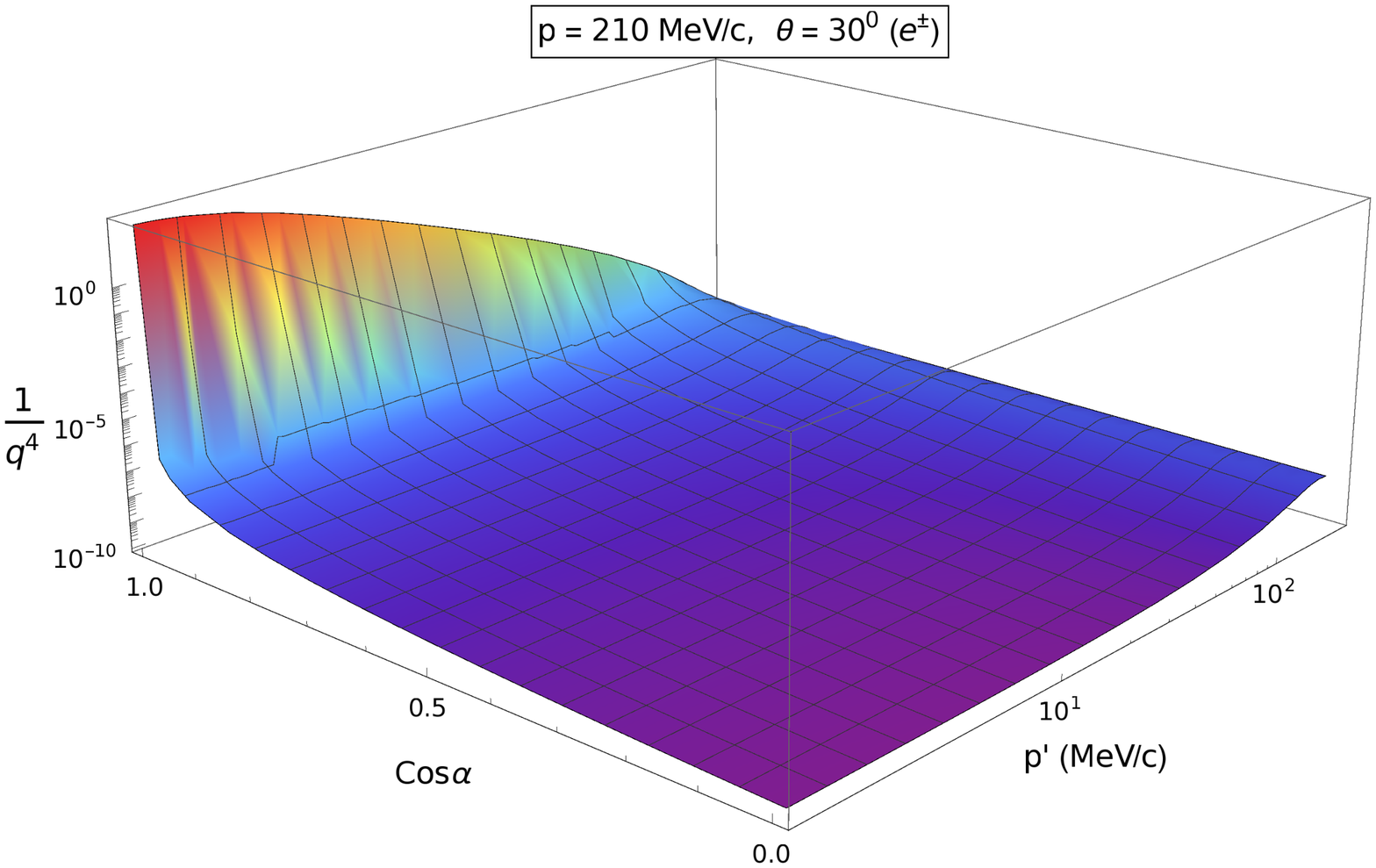} \qquad \includegraphics[scale=0.36]{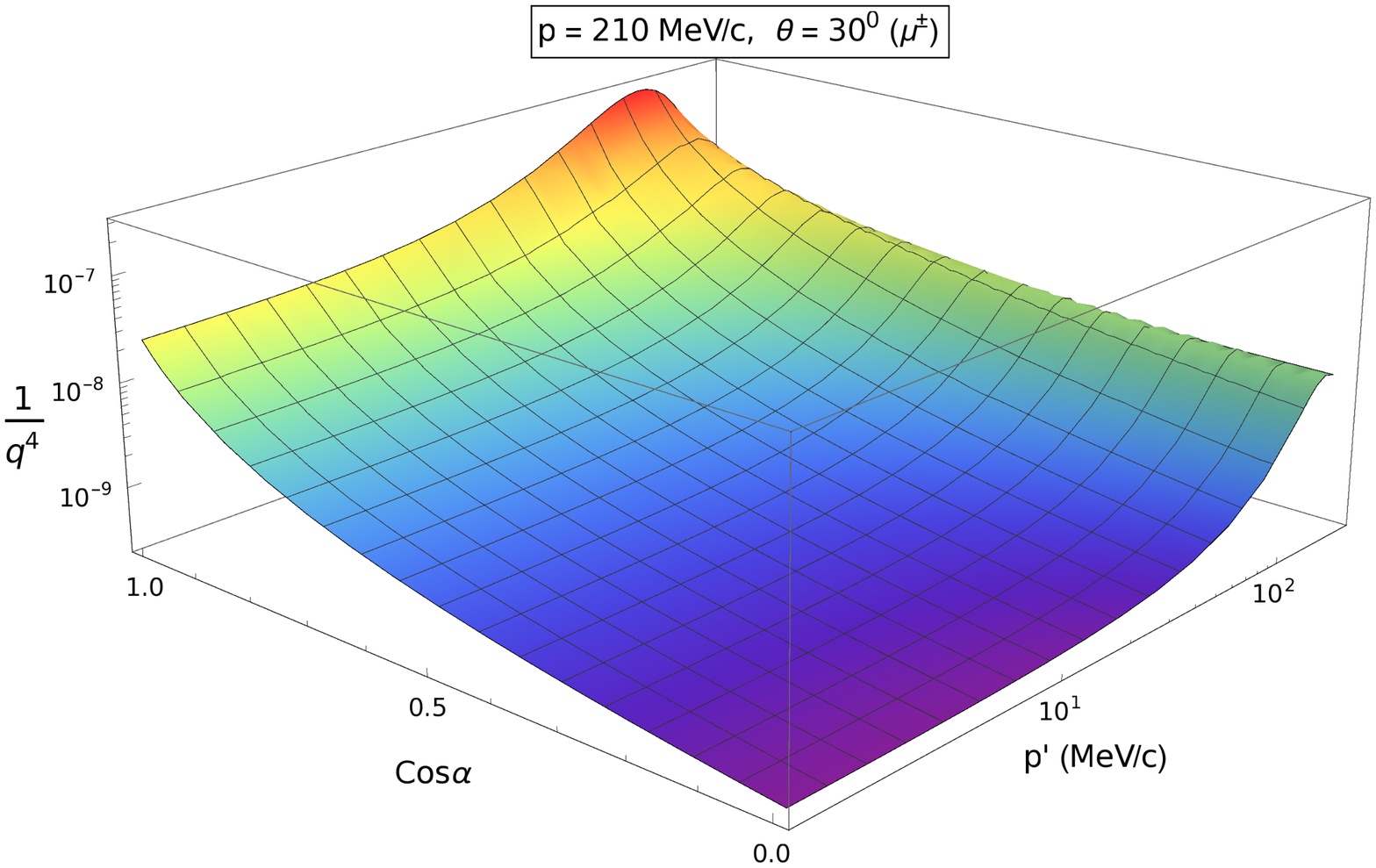} 
\caption{Behavior of $q^{-4}$ [in (MeV/c)$^{-4}$] as a simultaneous function of the outgoing lepton momenta 
$p^\prime$ and $0\leq\cos \alpha\leq 1$. The left plot is for the electron case and the right plot gives the 
muon results. In each case, the incident lepton momentum is $p=210$ MeV/c and the scattering angle is 
$\theta=30\degree$. In case of electron, there is a very large collinear enhancement for $p^\prime\to 0$ and 
$\alpha\to 0$ and manifests as a local maximum at small $p^\prime$ in Fig.~\ref{fig:tail}. This phenomenon 
is not noticeable in the muon spectrum. }
\label{fig:3D}
\end{figure*}

Second, we note that the bremsstrahlung cross section is directly proportional to $1/q^4$, Eq.~\eqref{eq:sigma5}. In 
Fig.~\ref{fig:3D} we display the behavior of $1/q^4$ as a simultaneous function of $p^\prime$ and $\cos\alpha$. 
In the  electron case, the figure clearly shows a large {\it collinear enhancement} of $1/q^4$ as 
$\alpha \to 0$ and $p^\prime$ is taken very small. This enhancement falls off sharply with the increasing values of both 
$\cos\alpha$ and $p^\prime$. In contrast, for the muons no such enhancement in $1/q^4$ is apparent for small $p^\prime$. 

%
%

\begin{figure}[tbp]
\centering
         \includegraphics[scale=0.32]{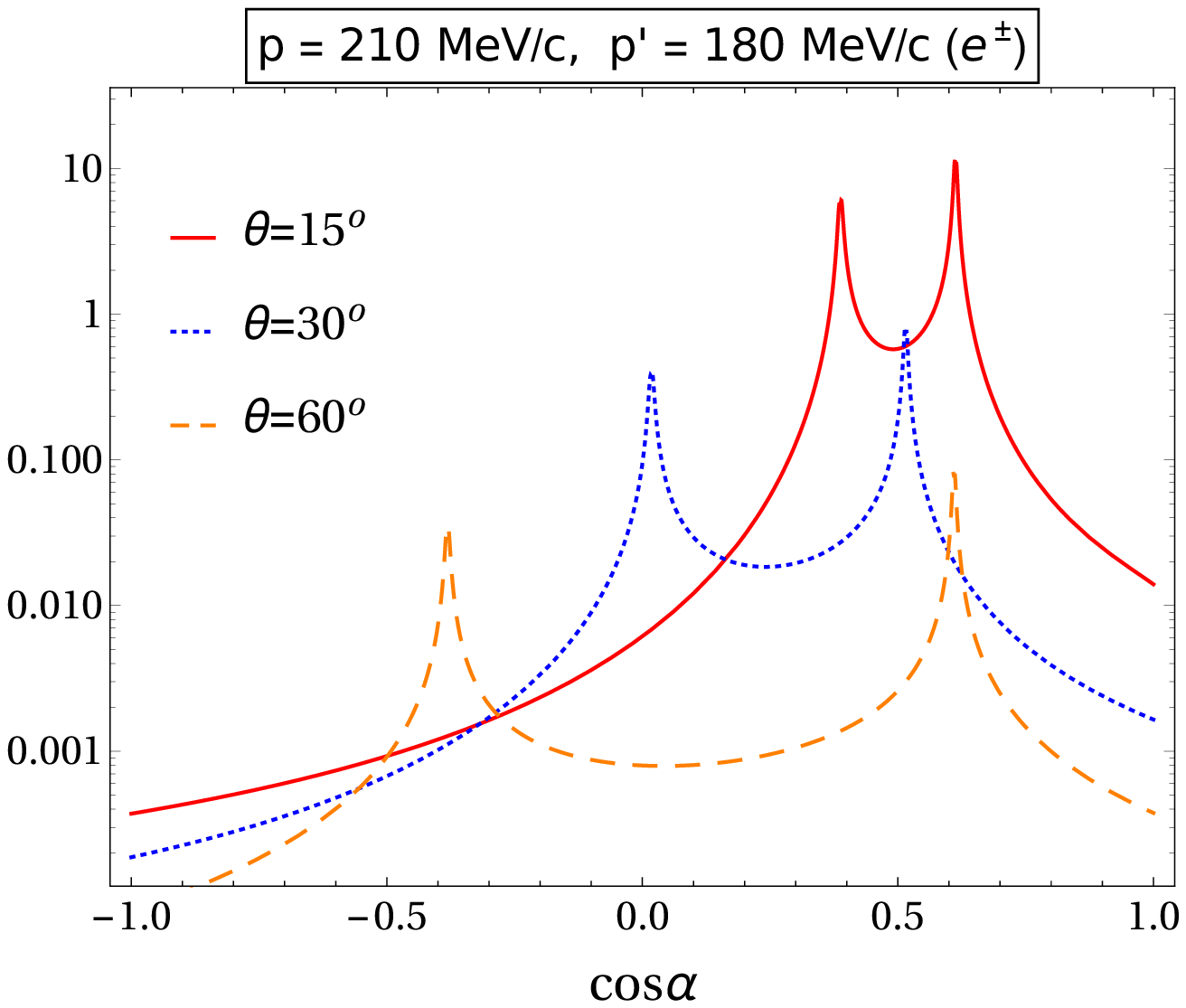} \includegraphics[scale=0.32]{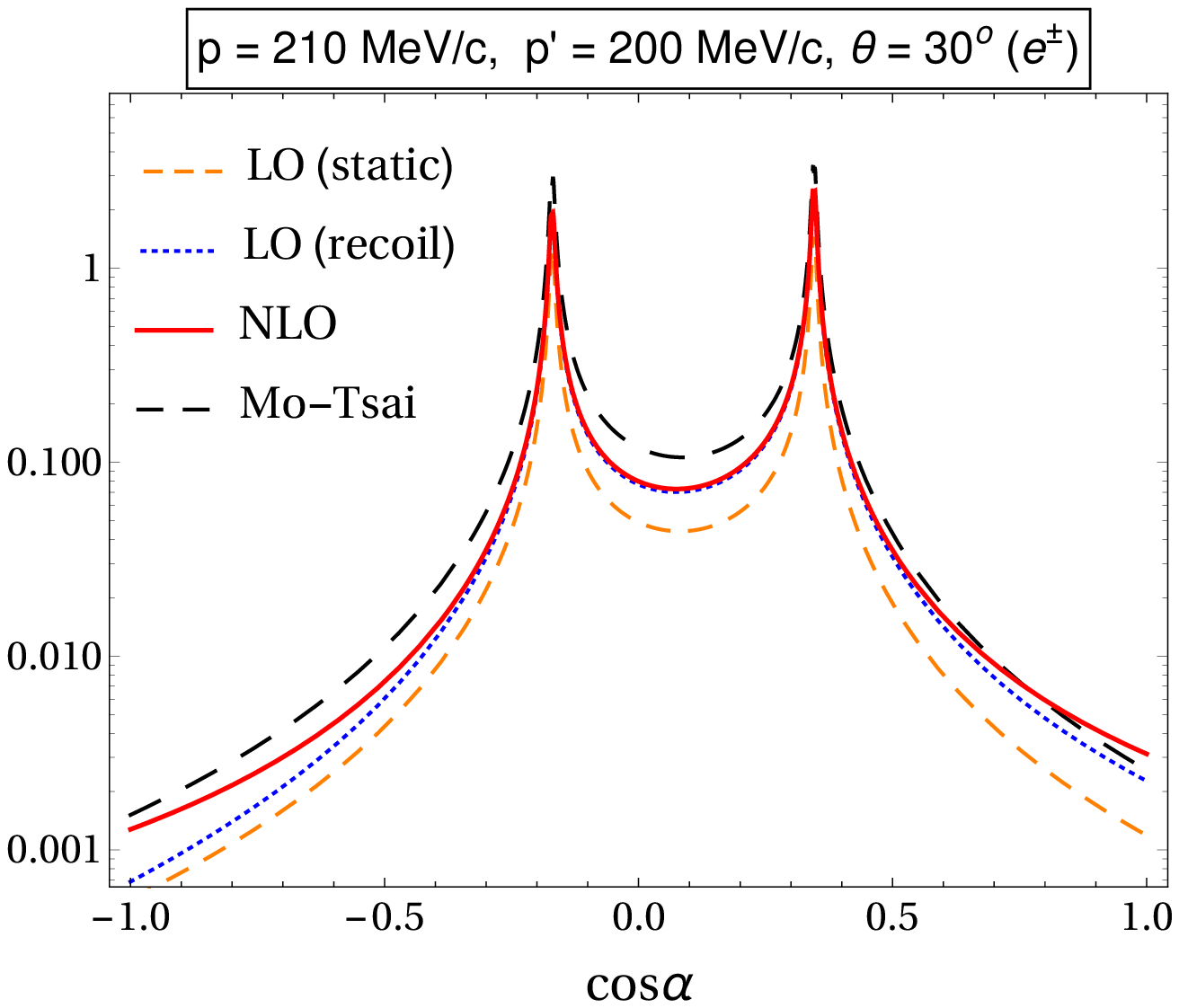}

\vspace{0.4cm}

         \includegraphics[scale=0.32]{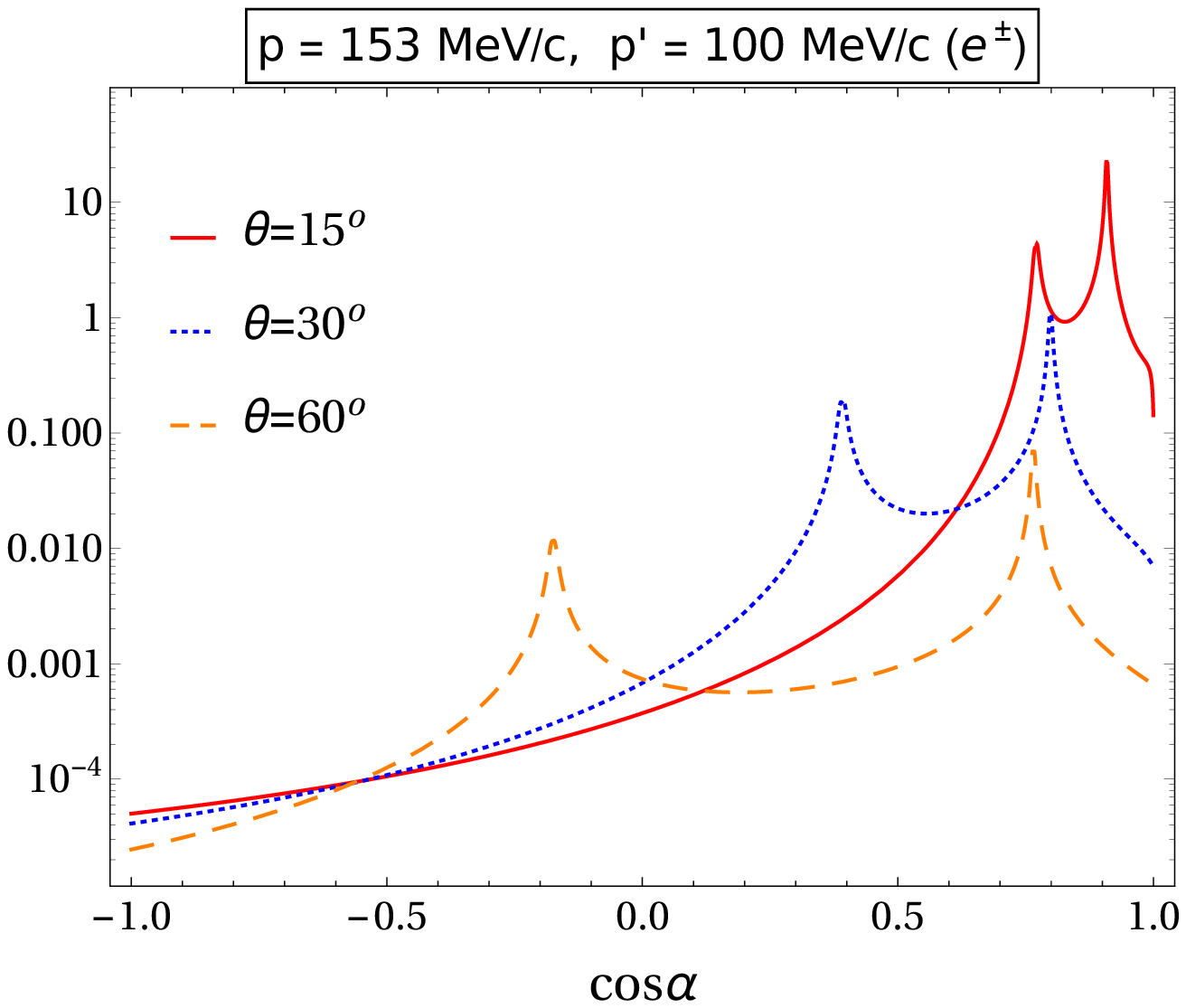} \includegraphics[scale=0.32]{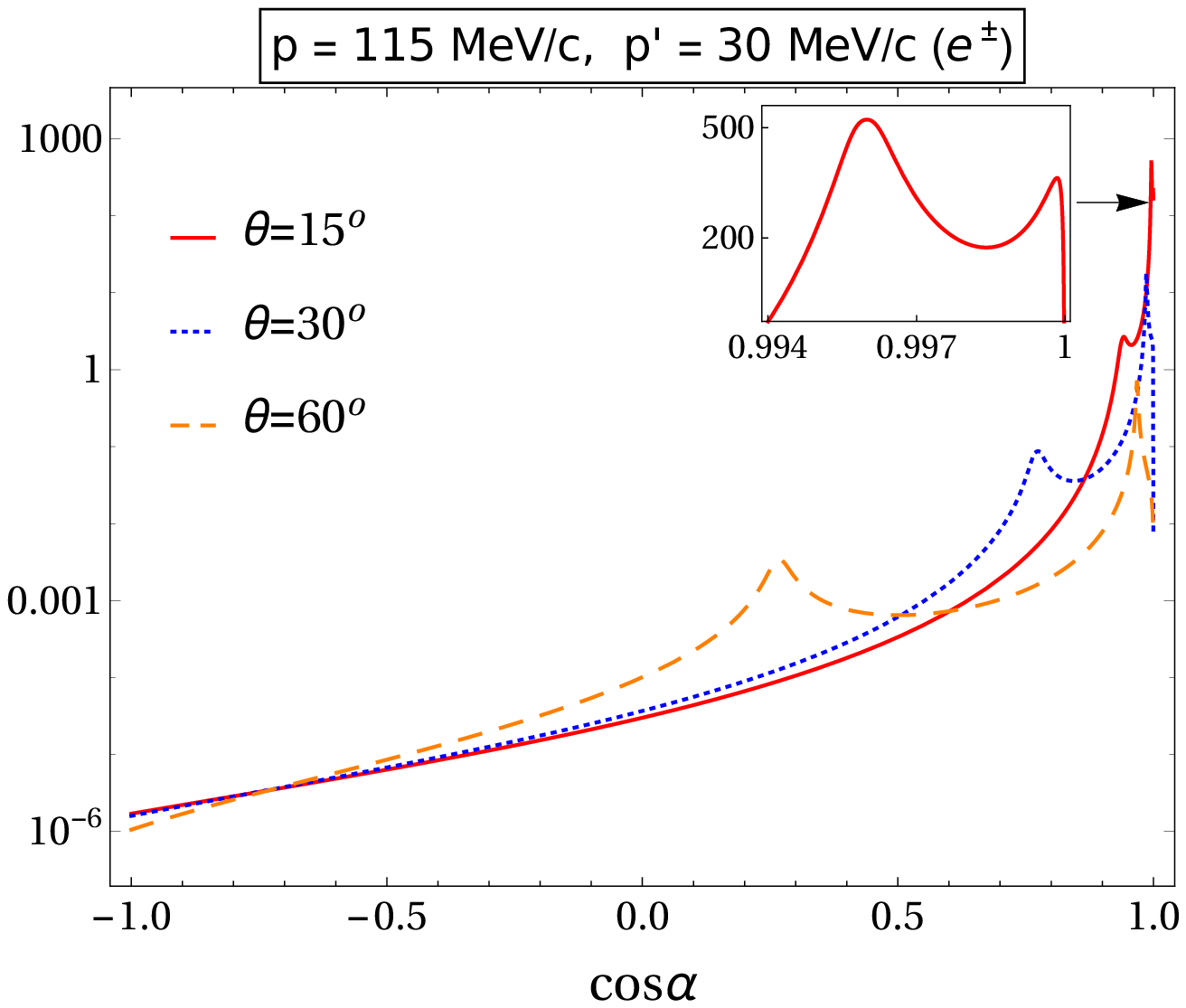}
\caption{The bremsstrahlung differential cross section up to and including NLO in $\chi$PT,
${\rm d}^3\sigma/({\rm d} |\vec{p}^{\, \prime}|\,{\rm d}\Omega^\prime_l\,{\rm d}{\rm cos}\alpha)$ (in mb/GeV/sr), versus 
$\cos\alpha$ for electron scattering for the three incident MUSE specified momenta $p=|\vec{p}\,|$, as displayed. For 
each $p$ just one value for the outgoing electron momentum $p^{\prime}=|\vec{p}^{\,\prime}|$ is plotted. In the two l.h.s. 
and bottom right plots, the solid (red) curves correspond to $\theta=15\degree$, the dotted (blue) curves to 
$\theta=30\degree$, and the dashed (orange) curves to $\theta=60\degree$. The insert in the lower right graph shows 
the dominant $\zeta$-peak and the additional third peak very close to $\cos\alpha = 1$. The top right graph compares 
the NLO result, Eq.(\ref{eq:sigma6}), with our LO evaluations, without [i.e., {\it static}, Eq.~(\ref{eq:sigma2})], and 
with the proton recoil terms of ${\mathcal O}(m^{-1}_p)$ in the phase space [i.e., {\it recoil}, Eq.~(\ref{eq:sigma5})]. 
In the same graph, the dashed curve shows the corresponding result obtained using the corrected expression for Eq.~(B.5) 
in Ref.~\cite{motsai1969} (see text). }
\label{fig:peak_e}
\end{figure}

In Fig.~\ref{fig:peak_e}, we display the result for the total differential cross section up to and including 
NLO in $\chi$PT, Eq.~(\ref{eq:sigma6}), versus the  cosine of the outgoing photon angle $\alpha$, for three MUSE 
specified incoming momenta, $p=|\vec{p}\,|=210,153,115$ MeV/c. For the bremsstrahlung process the outgoing 
lepton momentum can be chosen arbitrarily in the range $0\leq p^{\prime} \leq p^{\prime}_{elastic}(\theta)$, with 
$p_{elastic}(\theta) < p$ for a given scattering angle $\theta$. We only display the results for the MUSE specified 
kinematics with $p^{\prime}=30,100$ and $200$ MeV/c and for three forward angles: $\theta=15\degree,30\degree$ and 
$60\degree$. We also present a comparison of our NLO results with those of the LO (static and with recoil), as well 
as with the results obtained by using the corrected expression for Eq.~(B.5) in Ref.~\cite{motsai1969} (see clarification 
towards the end of this section). The differential cross section shows that the commonly used peaking 
approximation~\cite{tsai1961,motsai1969} is very well satisfied for the electron even at these low electron momenta $p$ 
and $p^\prime$. The double-peak structure is a distinctive feature of the angular radiative spectrum for ultra-relativistic 
particles.
%
%

The prominent double peaks occur for photon angle $\alpha$ close to the angle $\zeta$ (the $\zeta$-peak) for the 
incoming electron momentum, and the angle $\gamma$  (the $\gamma$-peak) for the outgoing electron momentum, as defined 
in Fig.~\ref{fig:kin_ref}. Moreover, it may be noted in the figure that for $\theta=15\degree$ for both the lower 
plots (as well as for $\theta=30\degree$ and $60\degree$ in the lower right plot) three peaks are generated with 
the $\zeta$-peak being the dominant one. In each case the rightmost peak-like structure very close to $\cos\alpha =1$, 
as shown by the insert plot in the lower right graph in Fig.~\ref{fig:peak_e}, can be attributed to the small $q^2$ 
(or alternatively, large $1/q^4$) behavior for angle $\alpha$ close to zero (also, see Fig.~11 in Ref.~\cite{motsai1969}). 
As expected from a classical bremsstrahlung angular spectrum (see, e.g., Ref.~\cite{motsai1969}), for relativistic 
electrons the emitted photons get collimated close to the incoming and outgoing directions of the electron momentum. 
Several further observations regarding the electron plots are in order:
\begin{itemize} 
\item The separation between the peaks increase with increasing scattering angle $\theta=\gamma-\zeta$. For 
$\theta\to 0$, the two peaks merge together into a single sharp peak, denoting collinear alignment of all  
momentum vectors. 
%
\item  For a finite electron mass $m_e$, incident momentum $|\vec{p}\, |$, and fixed angles ($\theta, \alpha$),  
the differential cross section becomes maximum when the outgoing three-momentum $p^{\prime}$ value corresponding to 
the minimum of $-q^2$ (maximum of $1/q^4$). 
\item For fixed scattering angle $\theta$ and three-momentum transfer $|\vec{Q} | = |\vec{p}-\vec{p}^{\,\prime} |$, 
the differential cross section decreases with increasing incident momentum $|\vec{p}\, |$.
\item For  fixed three-momenta $p$ and $p^\prime$, the differential cross section decreases with increasing 
scattering angle $\theta$.
\end{itemize} 
%
\begin{figure}[tbp]
\centering
         \includegraphics[scale=0.32]{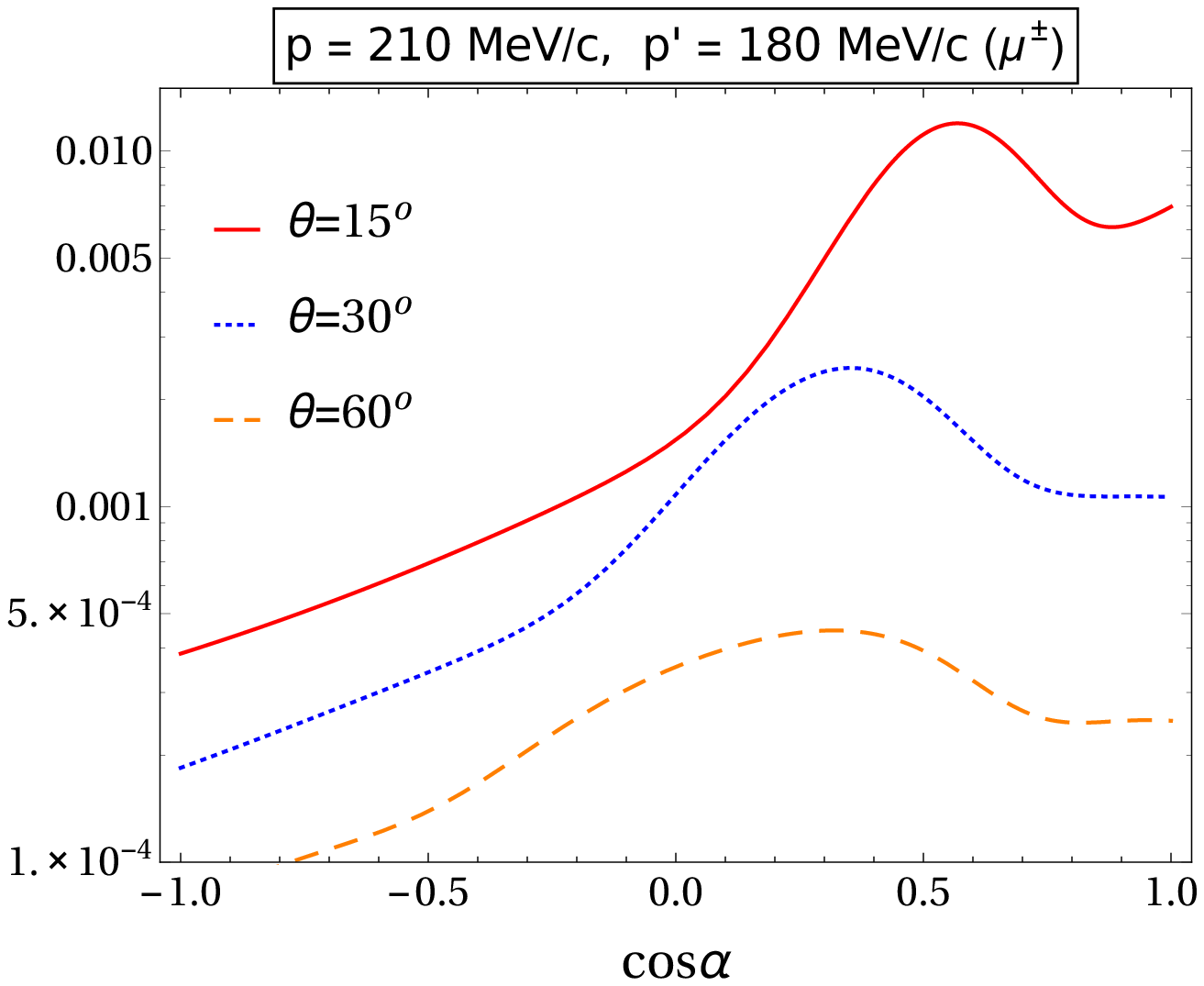} \includegraphics[scale=0.31]{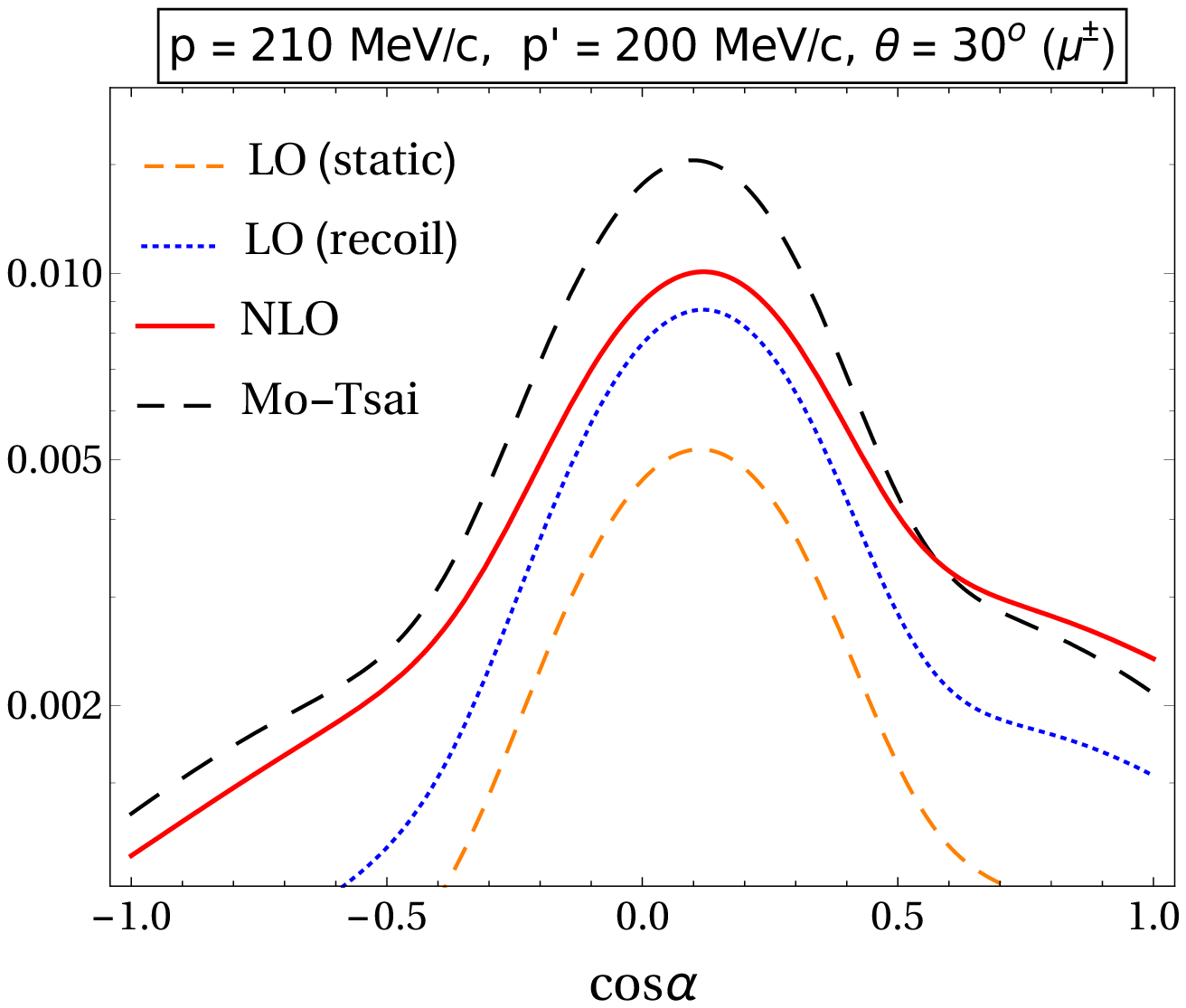}

\vspace{0.4cm}

         \includegraphics[scale=0.32]{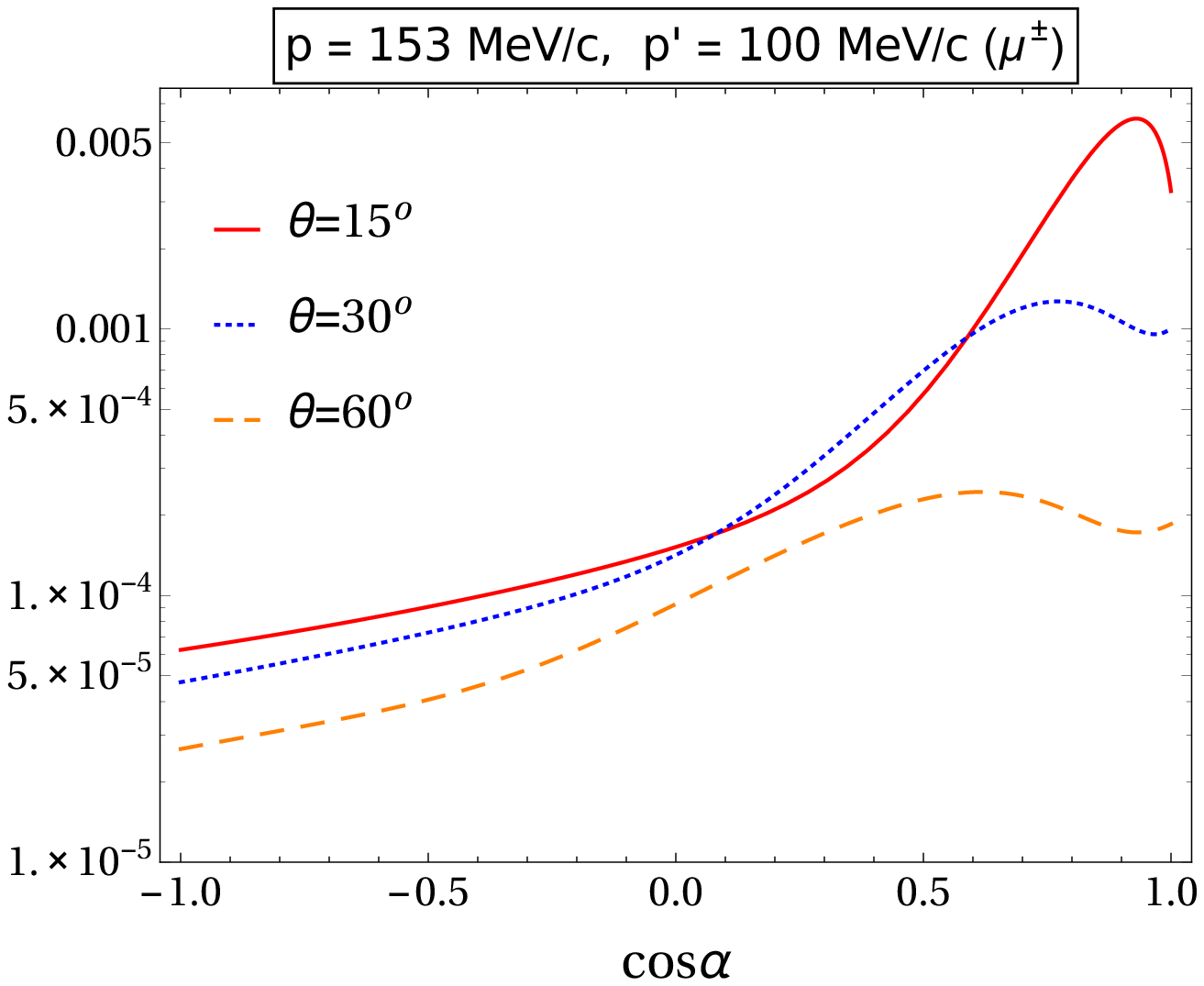} \includegraphics[scale=0.32]{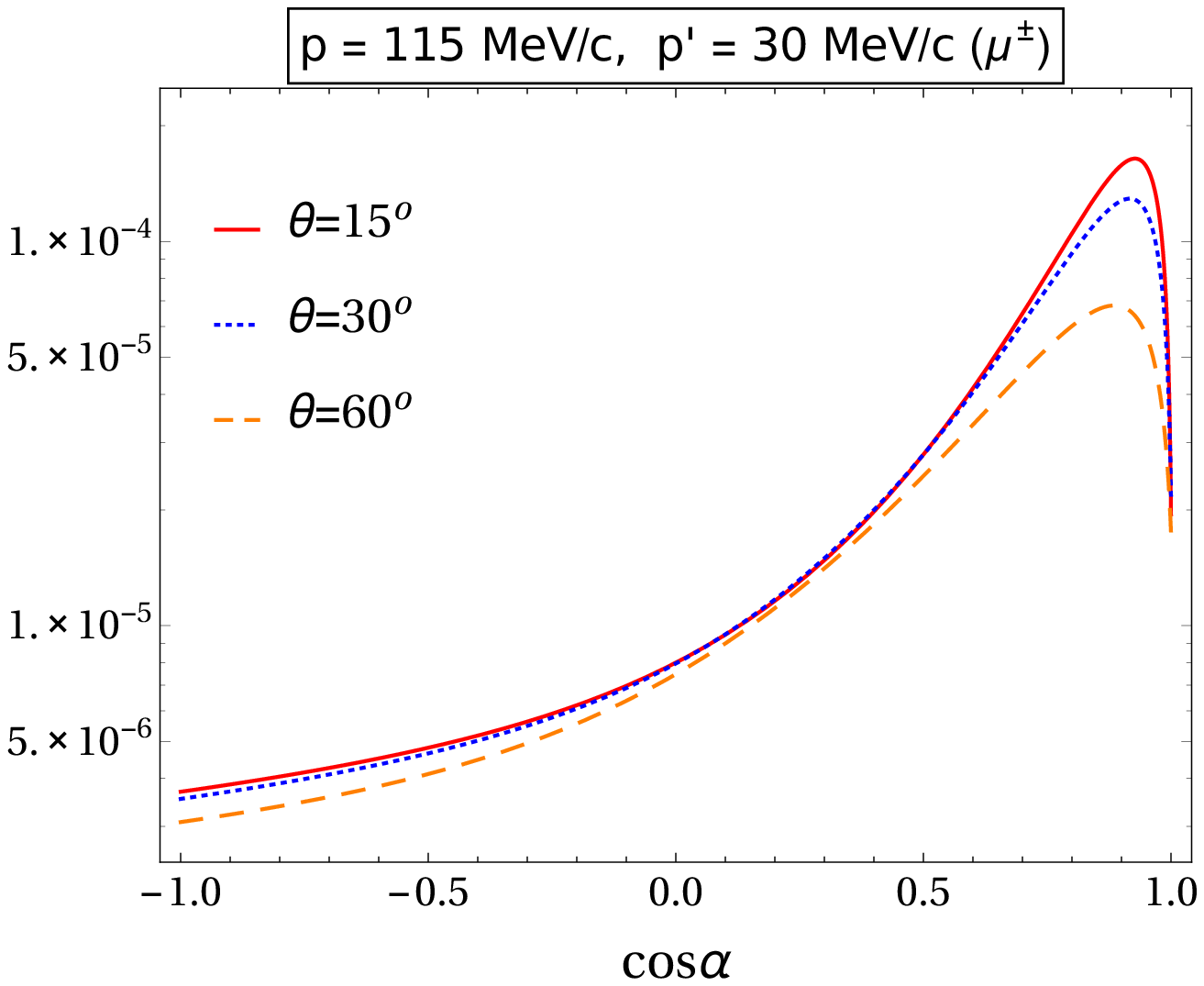}
\caption{ The  bremsstrahlung differential cross section up to and including NLO in $\chi$PT,
${\rm d}^3\sigma/({\rm d} |\vec{p}^{\, \prime}|\,{\rm d}\Omega^\prime_l\,{\rm d}{\rm cos}\alpha)$ (in mb/GeV/sr) 
versus $\cos\alpha$  for muon scattering for the three incident MUSE specified momenta $p=|\vec{p}\,|$, as displayed. 
For each $p$ just one value for the outgoing muon momentum $p^{\prime}=|\vec{p}^{\,\prime}|$ is plotted. See text 
and caption of Fig.~\ref{fig:peak_e} for details.}
\label{fig:peak_mu}
\end{figure}
%

In contrast to our electron scattering results, the $\cos\alpha$ dependence with the same kinematics for 
incoming muons is very different, as seen in Fig.~\ref{fig:peak_mu}. The initial muon momenta are not 
much larger than the muon mass, and the bremsstrahlung differential cross section versus $\cos\alpha$ has a 
broad angular spectrum. The plots in Figs.~\ref{fig:peak_e} and \ref{fig:peak_mu} clearly demonstrate that 
the so-called {\it peaking approximations}~\cite{motsai1969}, a widely used practical recipe for data analysis 
incorporating radiative corrections, while viable for electron scattering at the low-momentum MUSE kinematics, 
is not applicable for muon scattering at MUSE energies. As seen from Fig.~\ref{fig:peak_e} (top right plot), 
the expressions for the NLO ``interference'' corrections, namely, Eqs.~\eqref{eq:NLO_amps} - \eqref{eq:sigma7}, 
appear to yield contributions much smaller compared to the dominant LO result for the electron scattering. In 
contrast, for the muon scattering the NLO corrections are appreciably larger, as evidenced from the comparison 
plot in Fig.~\ref{fig:peak_mu}. Regarding our LO results, we have checked that the interference contribution in 
the LO differential bremsstrahlung cross section, Eq.~(\ref{eq:sigma5}), is much larger and dominates over the 
``direct'' contribution for both the electron and muon scattering (provided that the value for $p^\prime$ is not 
too small for the electron scattering case). As expected, the muon bremsstrahlung differential cross section is 
reduced by roughly two orders of magnitude compared to the corresponding electron cross sections for the same 
kinematic specification. 

\begin{figure}[t]
\centering
         \includegraphics[scale=0.29]{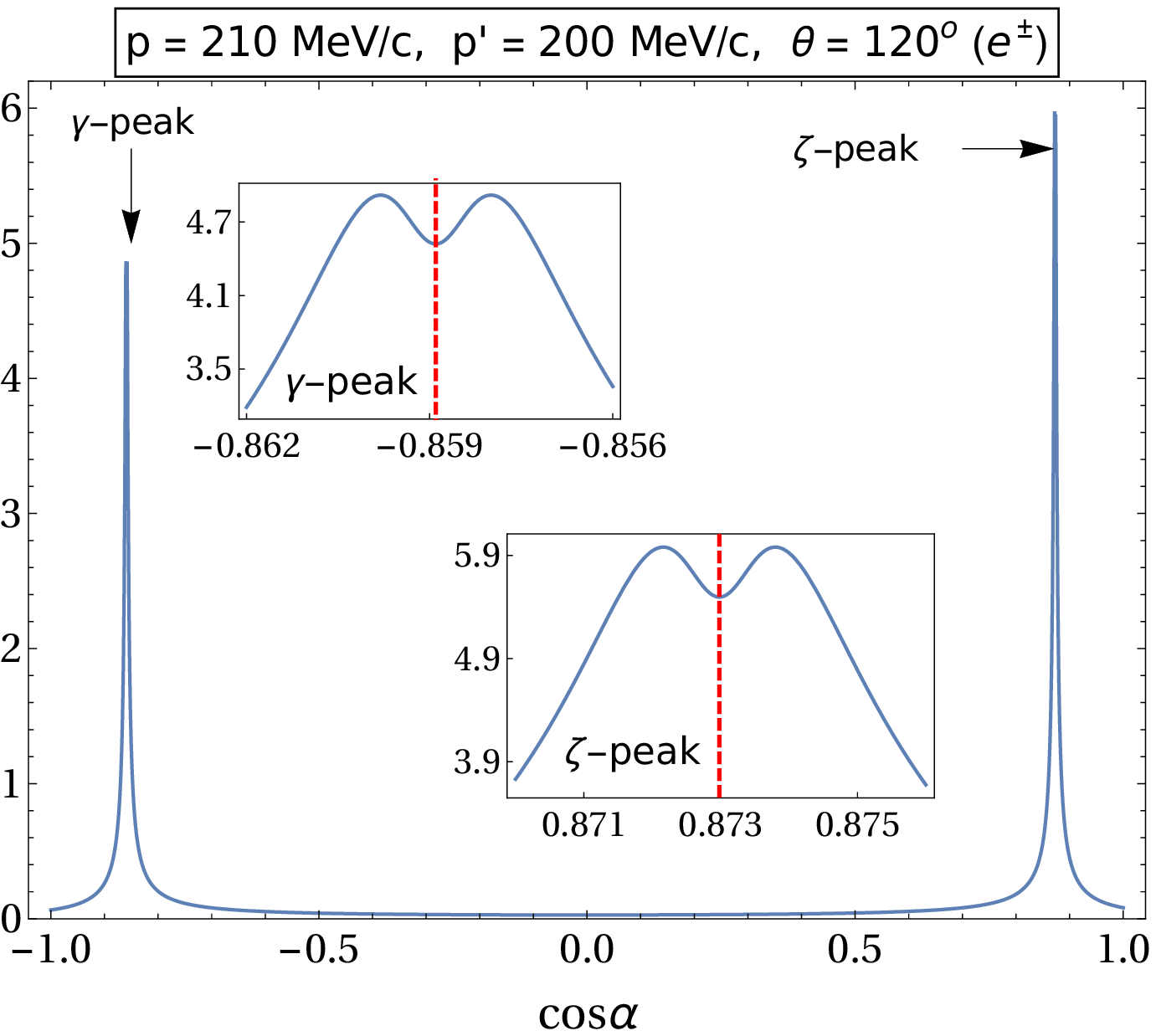} \includegraphics[scale=0.3]{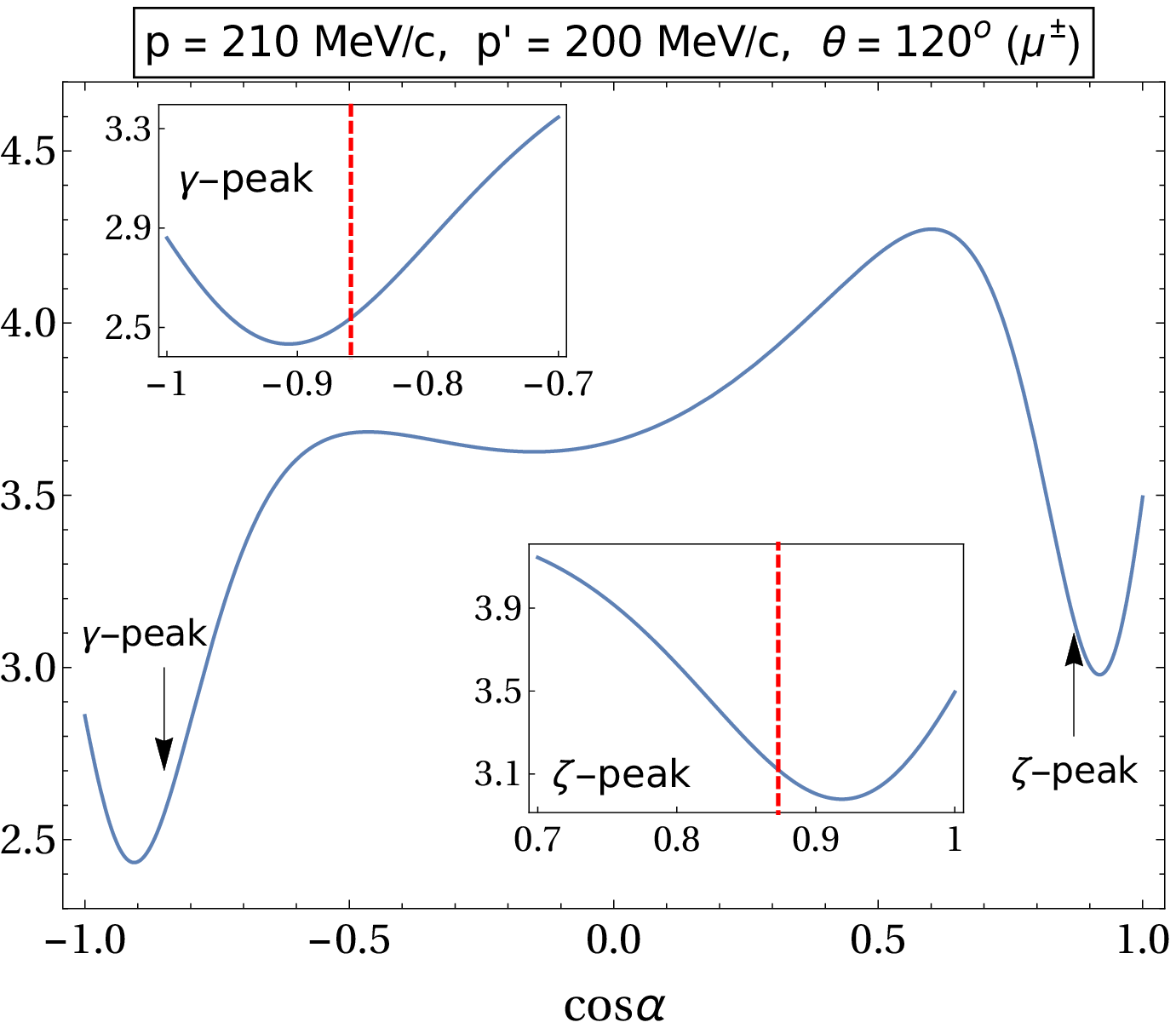}
\caption{ The bremsstrahlung differential cross sections up to and including NLO in $\chi$PT,
${\rm d}^3\sigma/({\rm d} |\vec{p}^{\, \prime}|\,{\rm d}\Omega^\prime_l\,{\rm d}{\rm cos}\alpha)$ (in mb/GeV/sr), versus 
$\cos\alpha$, indicating photon emissions from the leptons distributed within a shallow cone about their incident 
or scattered directions. The left plot is for electrons whereas the right plot is for muons. Both plots correspond to 
$p= 210$ MeV/c  and $p^\prime = 200$ MeV/c, and for the lepton scattering angle $\theta= 120\degree$. The red dashed 
lines indicate the directions of the incident and scattered leptons, i.e., $\cos\zeta$ and $\cos\gamma$, respectively. } 
\label{fig:cone}
\end{figure}

Zooming in on each peak in the electron angular $\cos\alpha$ dependence reveals the existence of a (3D) cone-like 
sub-structure, as displayed in Fig.~\ref{fig:cone}, i.e., the photon emission is (almost) collinear with the 
incoming and outgoing electron momenta. It may be recalled that for a charged relativistic particle with 
an acceleration parallel to its velocity $\vec{\beta}$, the  angular intensity distribution of the classical radiation 
corresponds to a cone with maximal opening angle $\sim{\mathcal O}(\sqrt{1-\beta^2})$ with respect to the direction of 
motion $\vec{\beta}$. The dashed vertical lines in Fig.~\ref{fig:cone} correspond in our reference system, 
Fig.~\ref{fig:kin_ref}, to the expected directions of the incoming and outgoing leptons. The effect of bremsstrahlung 
radiation results in the lepton recoiling away in a slightly different direction, leading to the characteristic cone-like 
feature for each peak with the vertex along the expected axis of the radiation cone. Unlike the $\cos\alpha$ dependence of 
the electron bremsstrahlung cross section, in case of the muon we observe significant interference effects between the two 
dominant broad angular peak structures arising entirely from the ``direct'' terms, labelled $[ \dots ]_{\rm dir(\gamma )}$ 
and $[ \dots ]_{\rm dir(\zeta )}$, and the ``interference'' contribution, labelled $[ \dots ]_{\rm int}$, in the LO 
expression Eq.~(\ref{eq:sigma5}). We find in addition a nominal NLO correction Eq.~(\ref{eq:sigma7}) to the 
$\cos\alpha$ behavior. These lead to the deviations of each minimum from the expected directions, as indicated by the 
vertically (red) dashed lines in the figure. Our graphic demonstrations support part of the analysis presented in 
Ref.~\cite{ent01} (e.g., see, Fig.~4 in this reference), where radiative corrections to $(e,e^{\prime}p)$ coincident 
experiments were discussed. It is to be, however, noted that if we were to reduce the value of the muon mass from 
its physical value, there should be a steady reduction of this observed mismatch between the vertical (red) dashed 
lines and the cone minima.
%
%

%
\begin{figure*}[t]
\centering
         \includegraphics[scale=0.33]{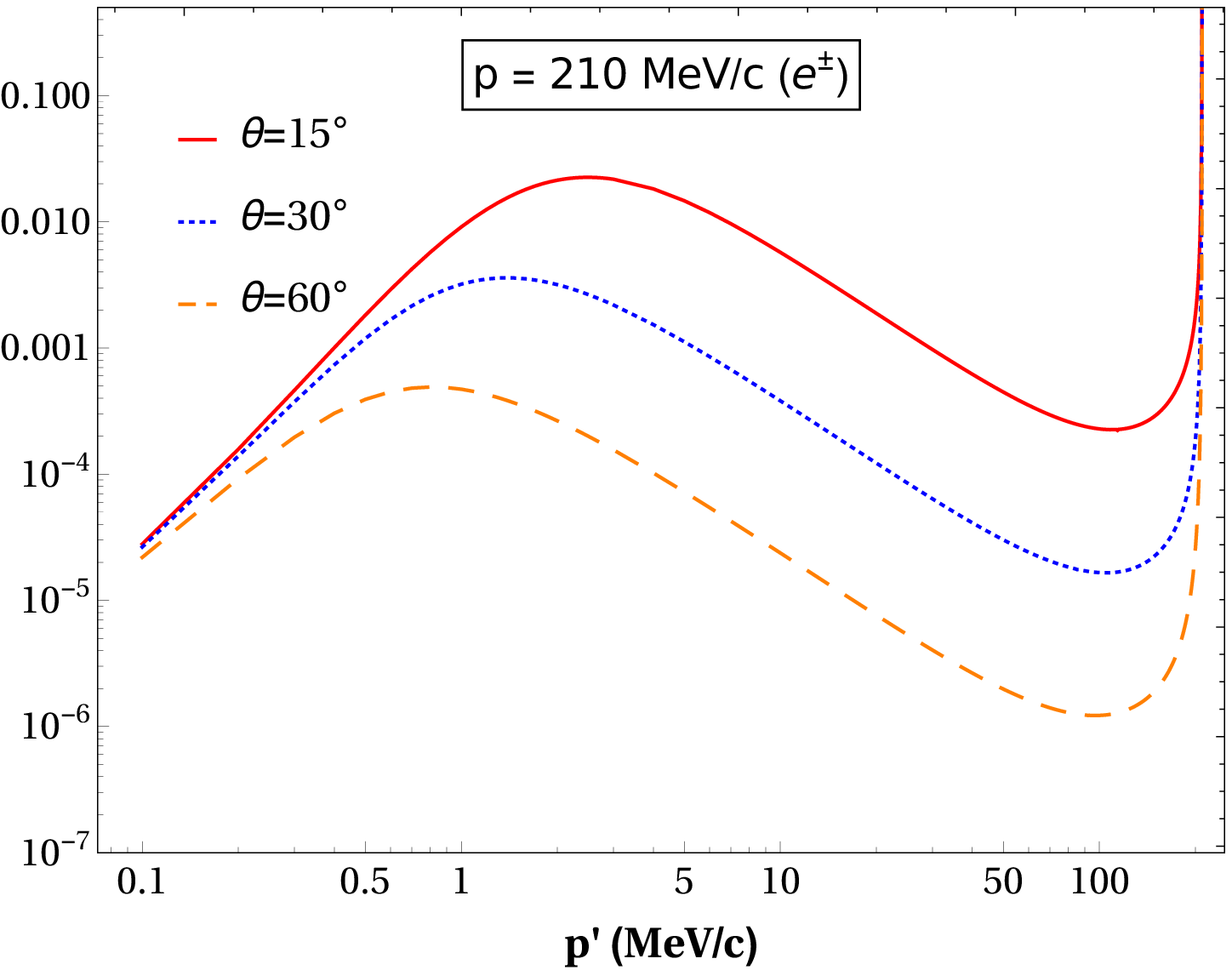} \qquad \includegraphics[scale=0.33]{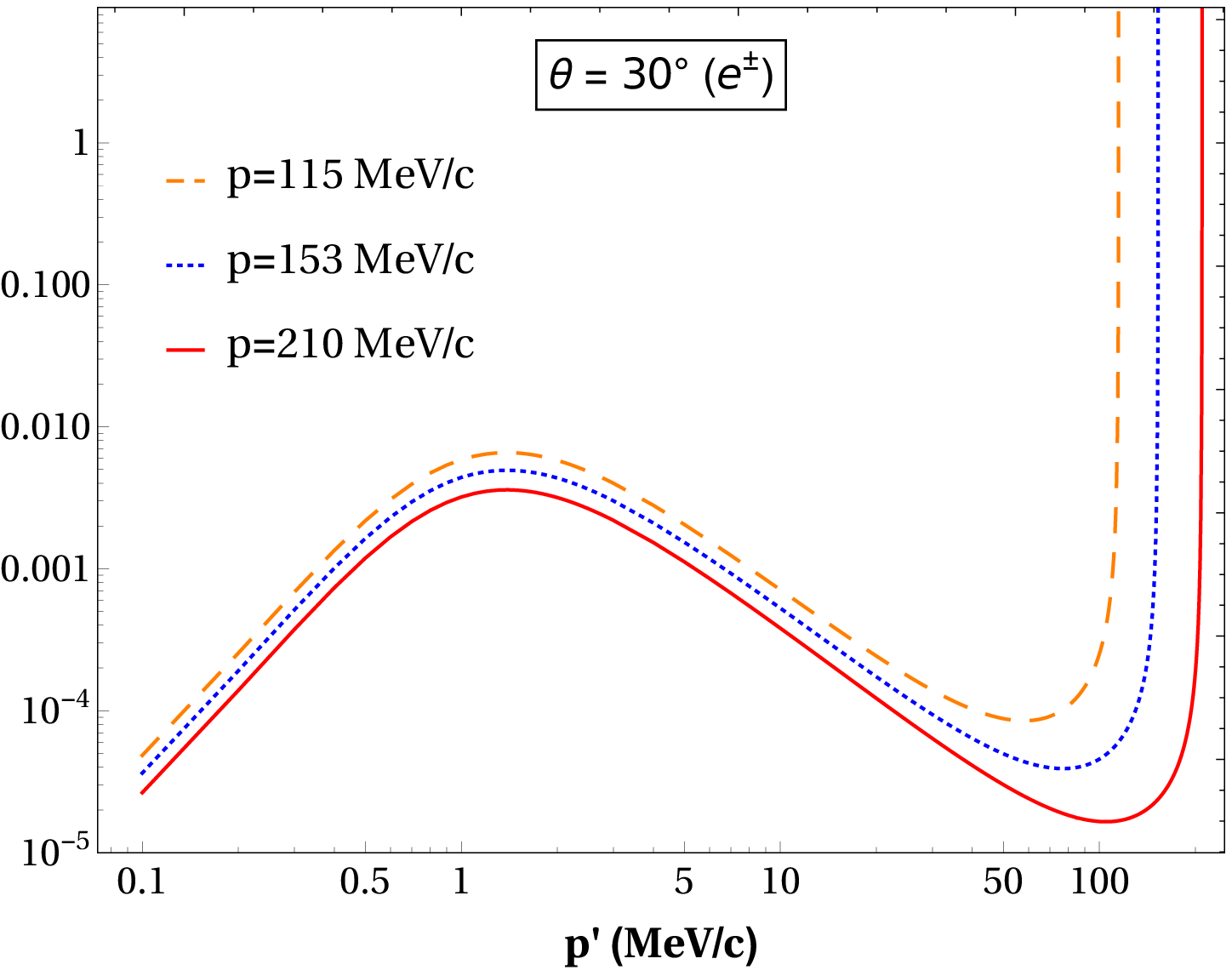} \qquad \includegraphics[scale=0.33]{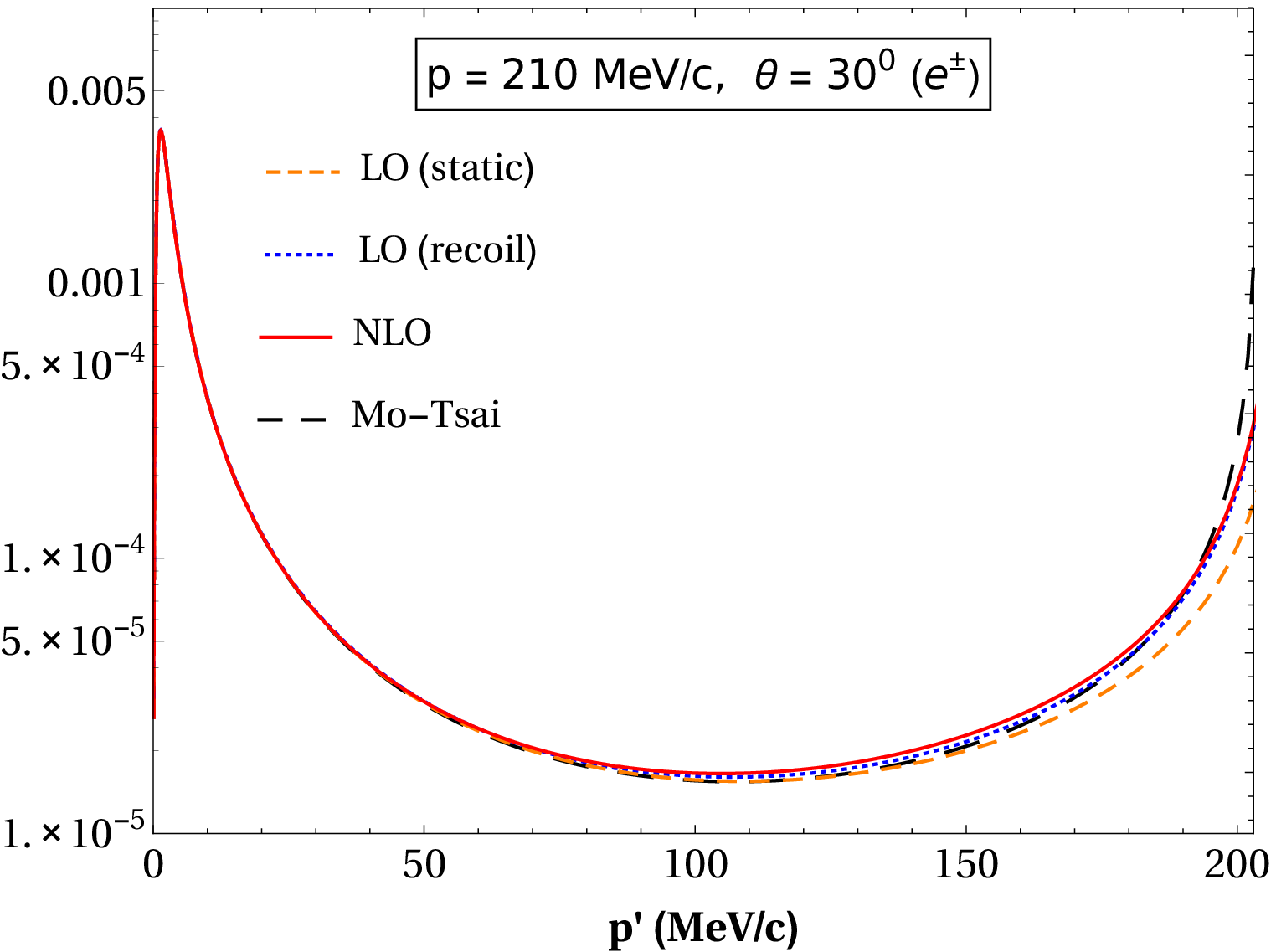}

\vspace{0.3cm}

         \includegraphics[scale=0.33]{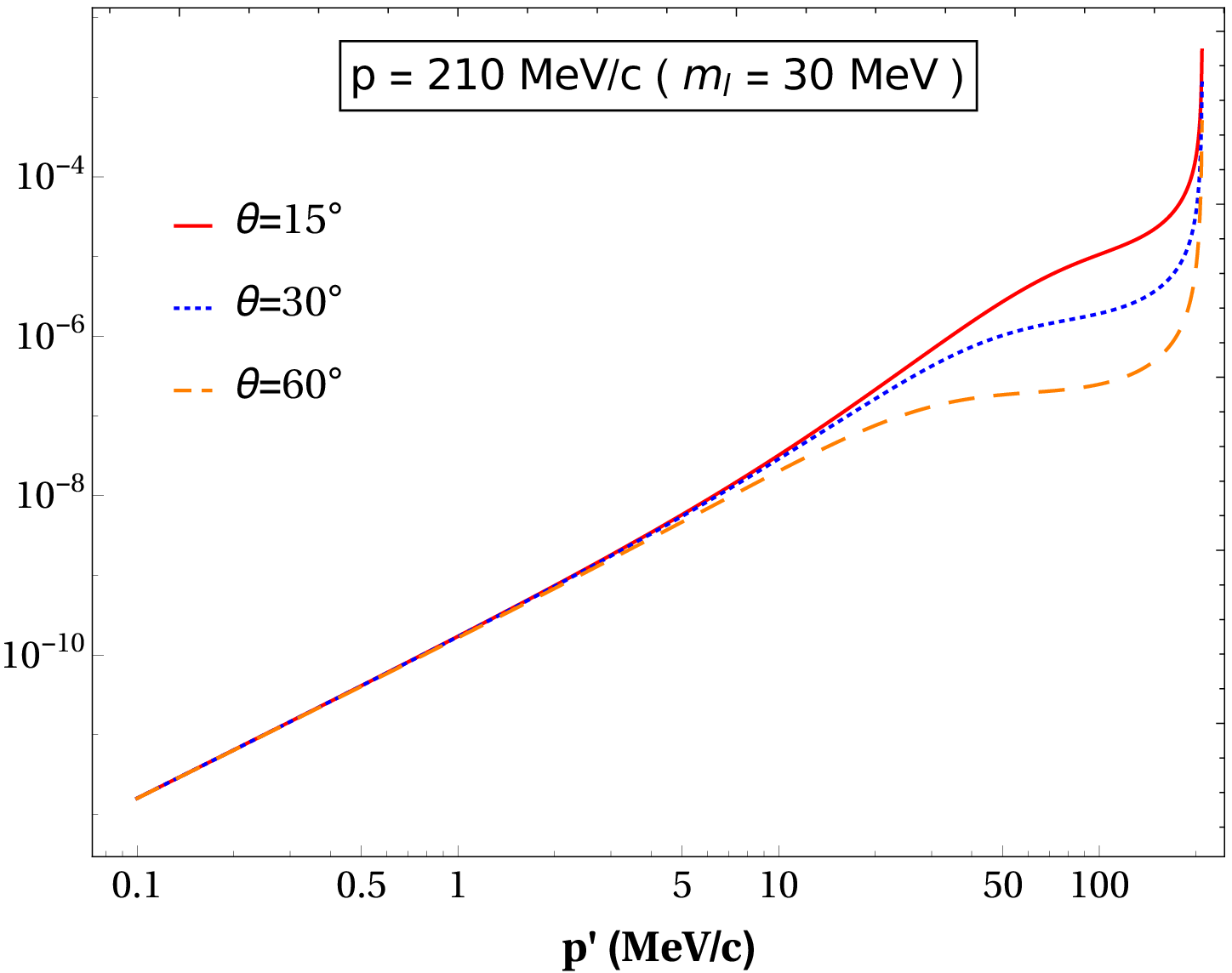} \qquad \includegraphics[scale=0.33]{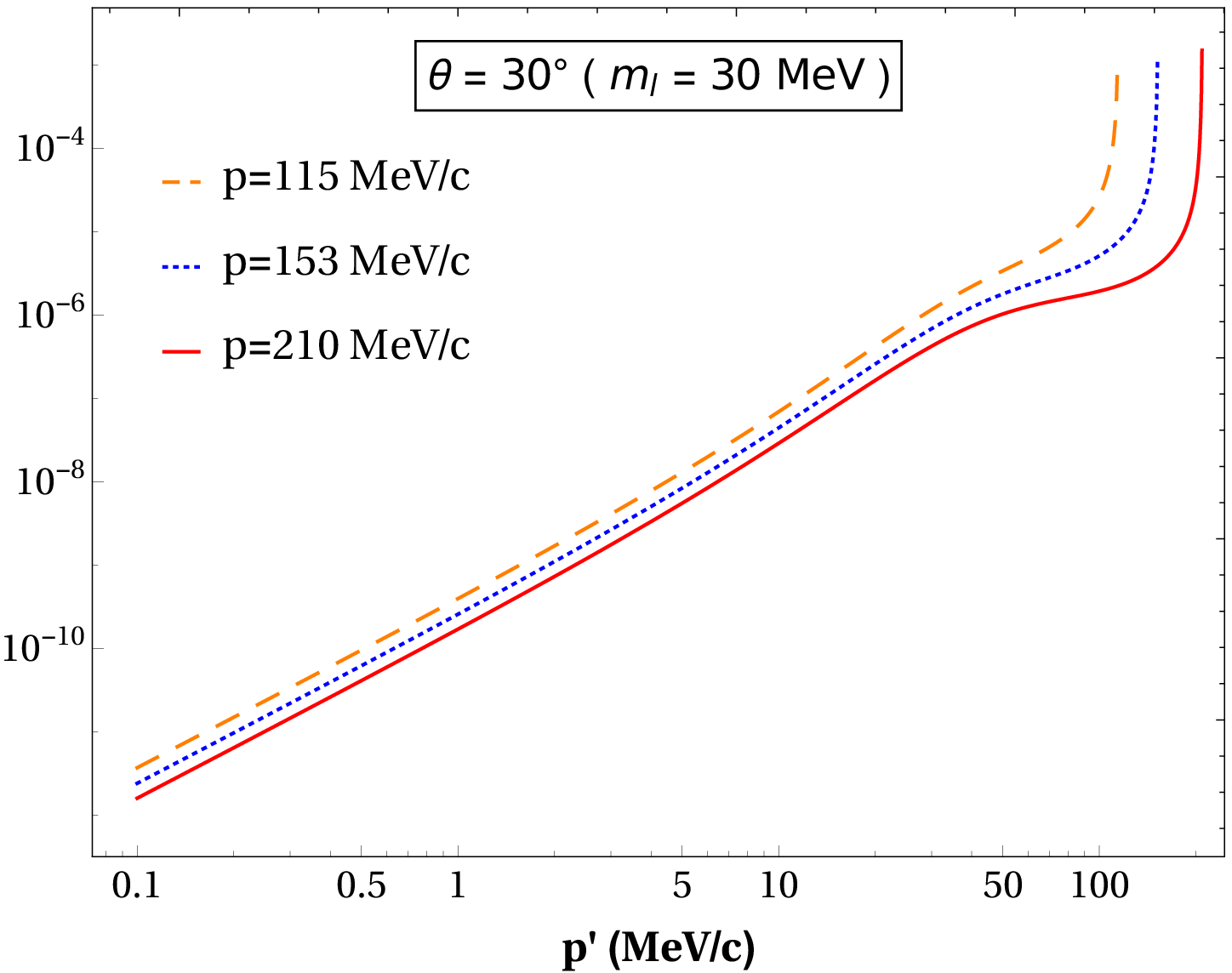} \qquad \includegraphics[scale=0.33]{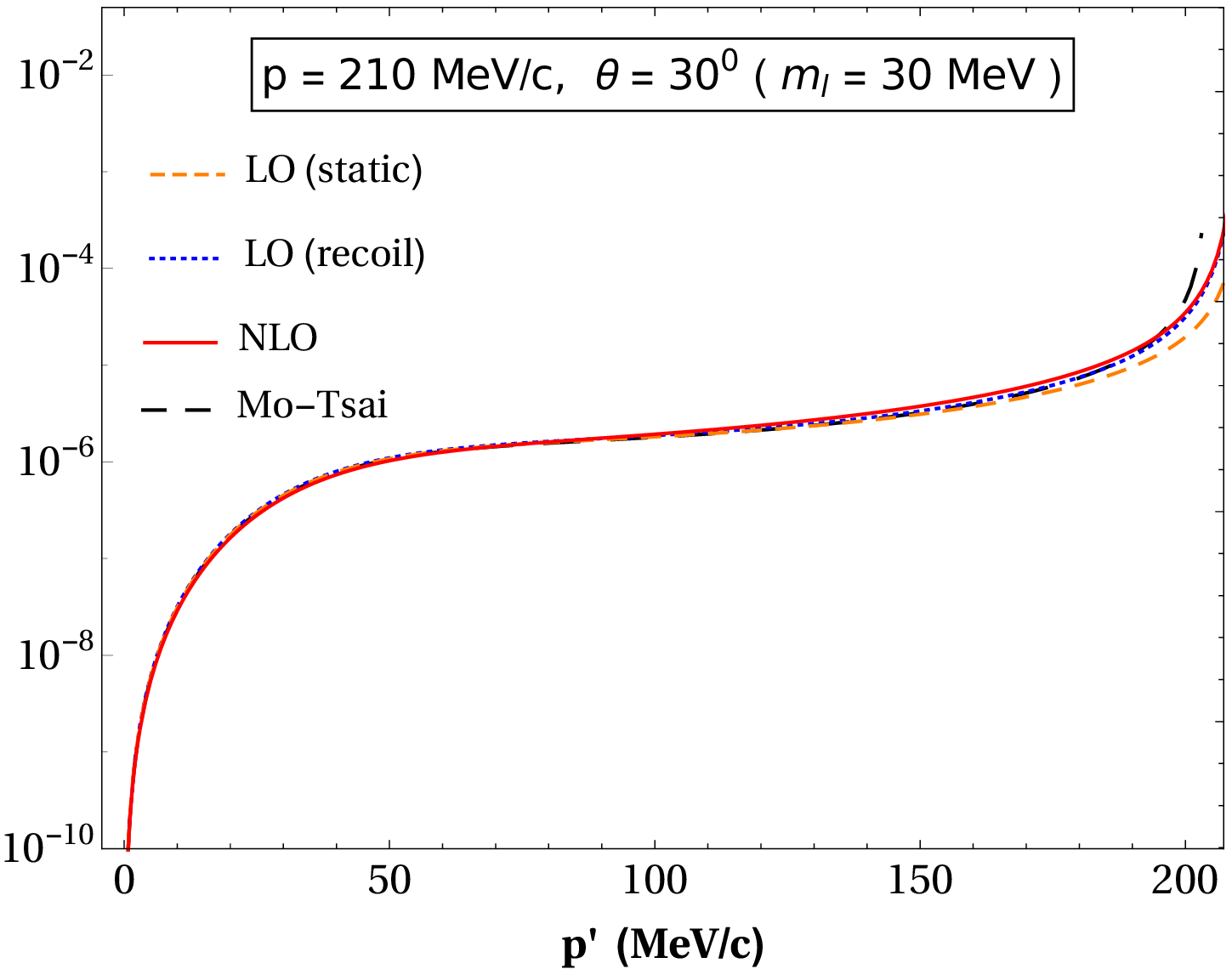}

\vspace{0.3cm}

         \includegraphics[scale=0.33]{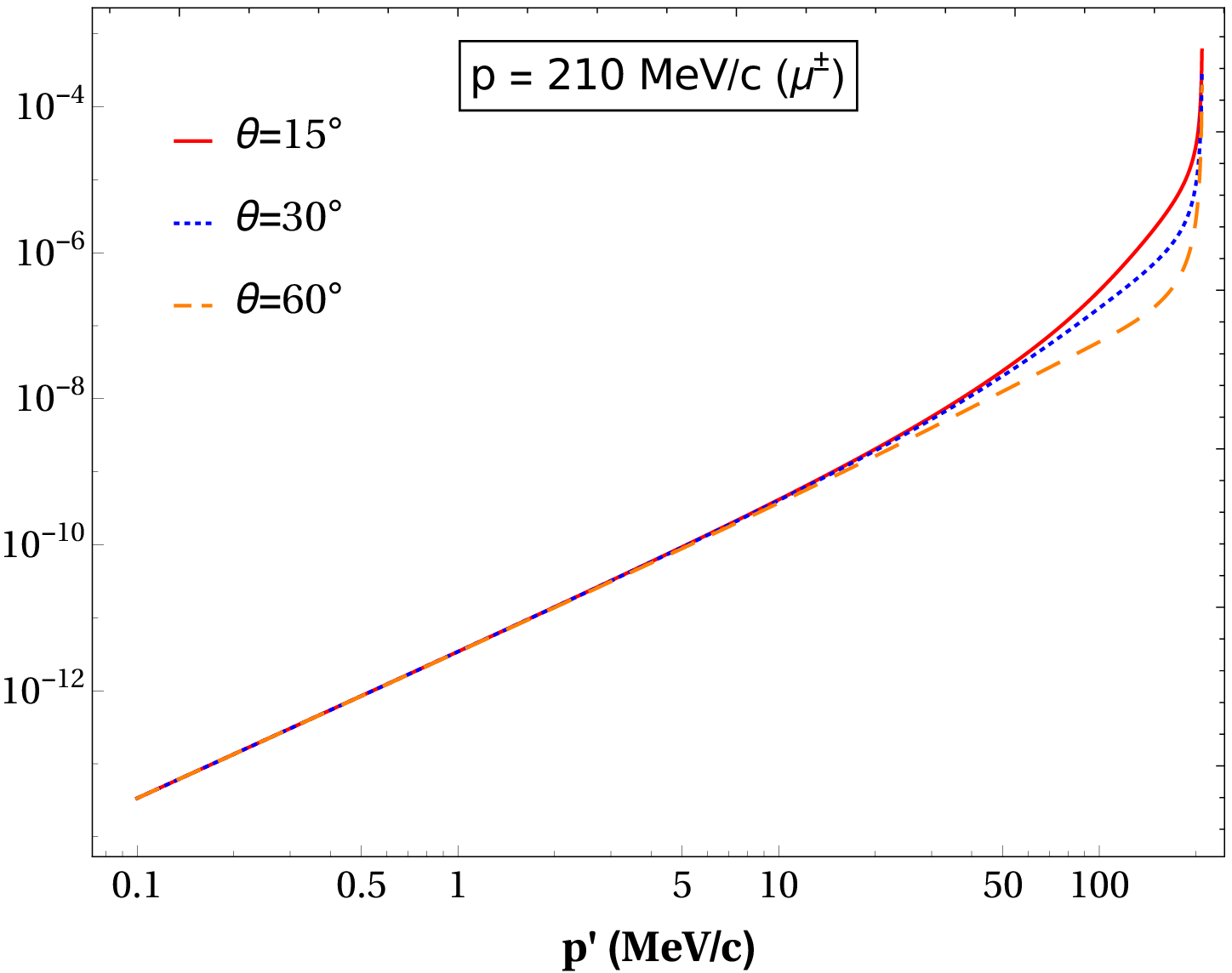} \qquad \includegraphics[scale=0.33]{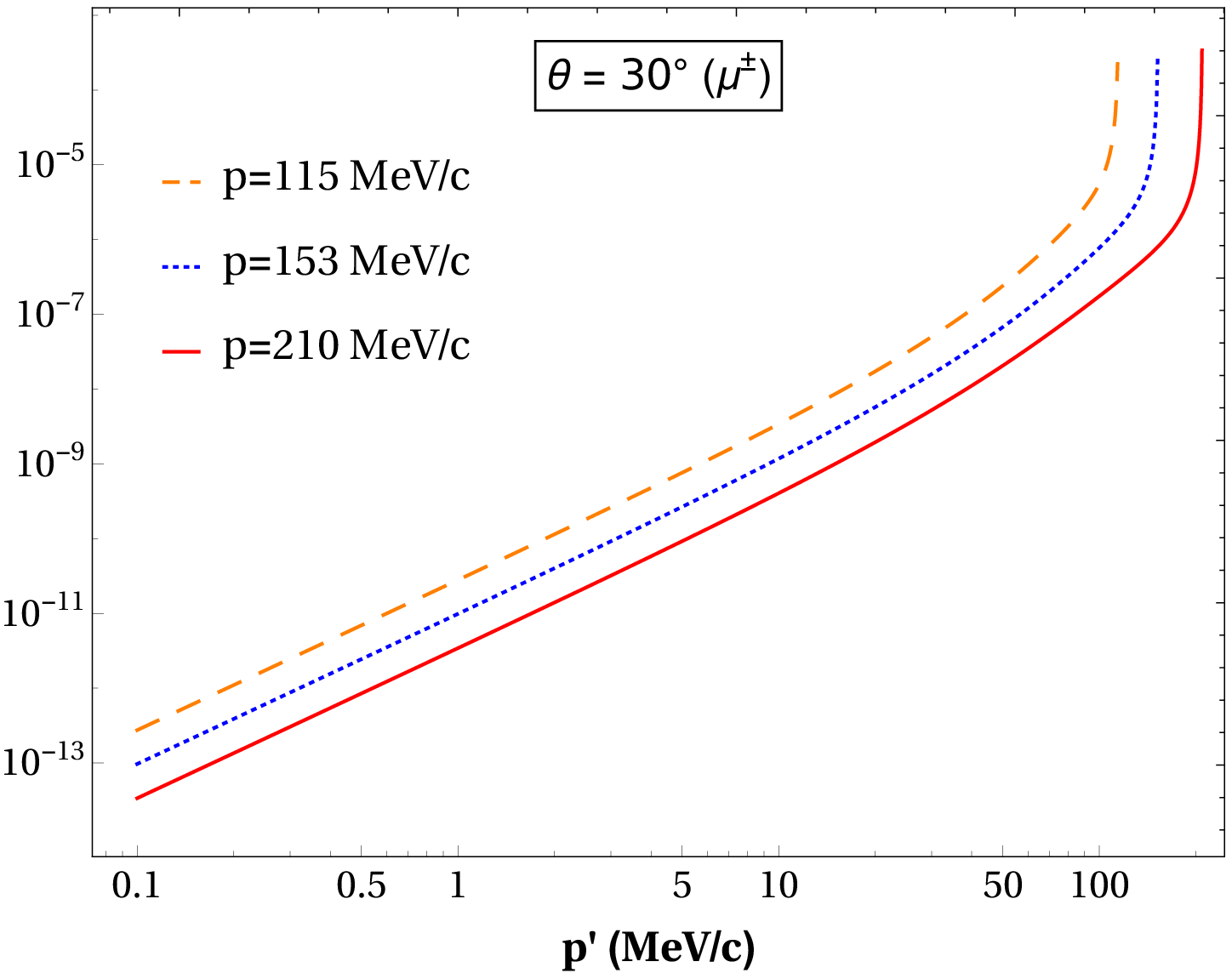} \qquad \includegraphics[scale=0.33]{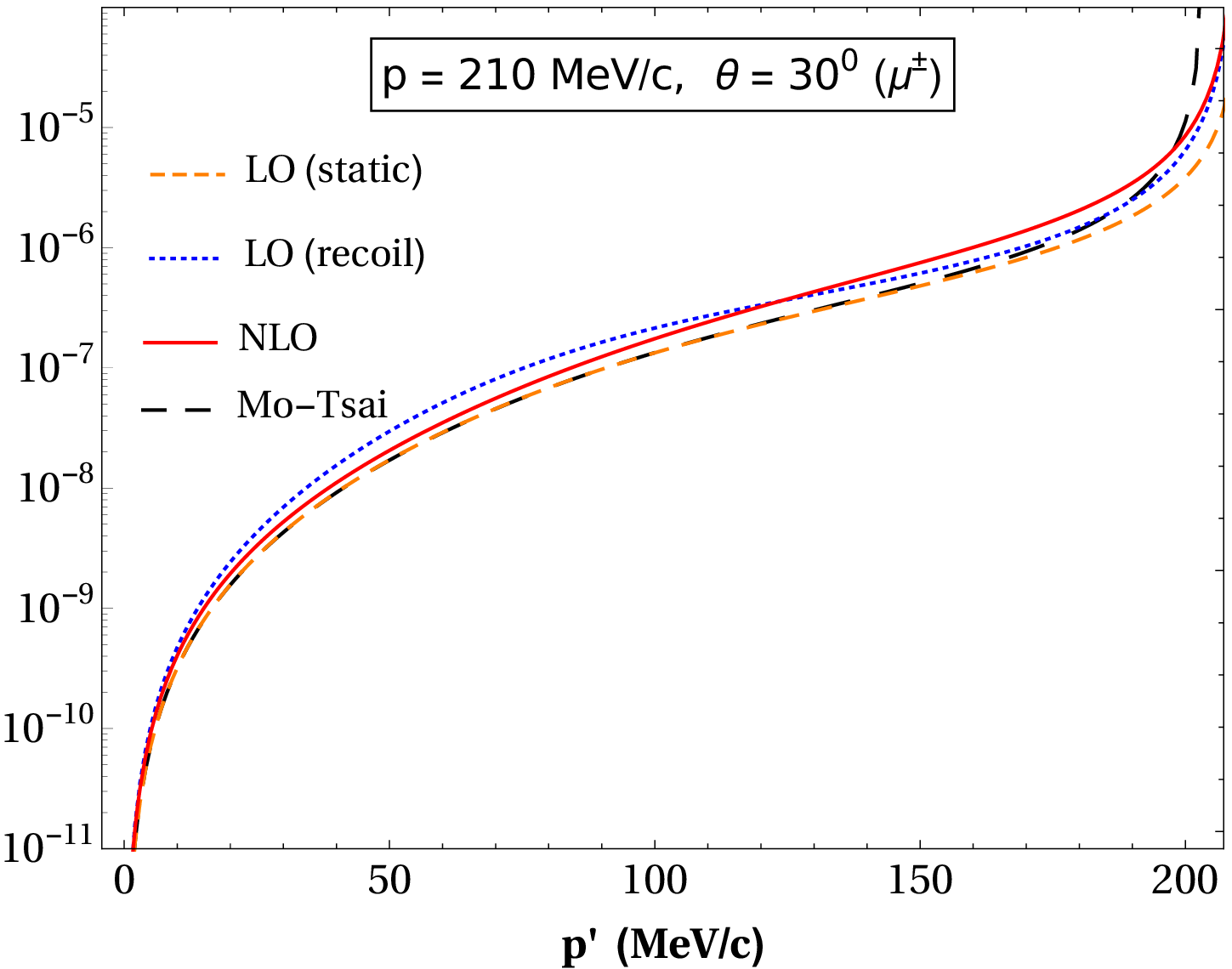} 
\caption{The ``radiative tail spectrum" cross section up to and including NLO in $\chi$PT,
${\rm d}^2\sigma/({\rm d} |\vec{p}^{\, \prime}|\,{\rm d}\Omega^\prime_l)$ (in mb/GeV/sr) is plotted as a function of 
the scattered lepton momentum $|\vec{p}^{\,\prime}|$, for incoming lepton momenta $|\vec{p}|$ and scattering angles $\theta$ 
specified by  MUSE. The plots in the upper (lower) row correspond to electron (muon) scattering, while the plots in the middle 
row correspond to the results for an intermediate lepton mass, $m_l\sim 30$ MeV. In the left column plots, 
where $|\vec{p}\,|=210$ MeV/c, we display the cross sections for $\theta=15\degree$ solid (red) curve, 30$\degree$ 
dotted (blue) curve, and 60$\degree$ dashed (orange) curve. For the middle column plots, we show the cross sections 
with $\theta=30\degree$ for the three incoming MUSE specified momenta, $p=210$ (solid), $153$ (dotted) and $115$ (dashed) MeV/c 
in the static approximation $m_p\to\infty$. The right column plots (with non-log scale abscissa) 
compare our NLO results with the {\it static} and {\it recoil} LO evaluations, Eqs.~(\ref{eq:sigma2}) and (\ref{eq:sigma5}), 
respectively, as well as with the results obtained using the corrected expression for Eq.~(B.5) in Ref.~\cite{motsai1969}, 
but ignoring the proton form factors. }
\label{fig:tail}
\end{figure*}  
%
The NLO results for the electron and muon bremsstrahlung ``radiative tail" cross section given in Eq.~(\ref{eq:sigma6}) 
for the MUSE specified values of the incoming lepton momenta, i.e., $p=|\vec{p}\, |=210, 153$ and $115$ MeV/c, are 
displayed in Figs.~\ref{fig:tail}. We only display the plots for the forward scattering angles, $\theta=15\degree,30\degree$, 
and $60\degree$, also specified for MUSE measurements. The bremsstrahlung cross section  versus $p^\prime$ is plotted from 
0.1 MeV/c up to $p^\prime_{max} = p^\prime_{elastic} - \Delta p^\prime$, where we have chosen $\Delta p^\prime = 0.1$ MeV/c in order 
to avoid the IR singular region. There is clearly a large IR enhancement of all the plots toward the large $p^\prime$ endpoint 
region. Without a proper treatment of the radiative IR divergences, our large $p^\prime$ results are beset with large 
uncertainties, though in the low-momentum region $p^\prime \lesssim 100$~MeV/c (suitable for MUSE) our results are reasonably 
accurate. We note that as $p^\prime$ tends toward zero, the differential cross section also goes  to zero for the both muon 
and electron cases. However, for the electron tail spectrum the cross section reaches a local maximum before going to zero 
as $p^\prime\to 0$.

This local maximum at small $p'$ values is primarily due to the dominant behavior of our static LO radiative tail cross 
section. When the outgoing electron momentum $\vec{p}^{\,\prime}\to 0$, the photon emission from the electron becomes 
(almost) collinear with the direction of the incident electron momentum $\vec{p}$ (reflected in the peaking approximation 
in Figs.~\ref{fig:peak_e} and \ref{fig:cone}) and  the momentum transfer, $\vec{q}=\vec{p} - \vec{p}^{\,\prime} - \vec{k}$. 
If we artificially let the electron mass go towards zero, we will encounter a {\it mass singularity} when $p^\prime=0$. The 
small electron mass effectively regularizes this singularity, and a remnant of this divergence manifests itself 
as a local maximum in the cross section. In that case the cross section goes to zero as $p^\prime\to 0$. 
%
%

Another way to examine the gradual development of the local maximum in the electron cross section is to consider our 
muon results presented in Fig.~\ref{fig:tail} (lower row plots) and artificially lower the value of the muon mass from 
the physical value, i.e., $m_\mu=105.7$ MeV. We find that for a lepton mass, $m_l\sim 30-40$ MeV, the cross section starts 
developing a ``shoulder", which for still smaller lepton mass values develops into a local maximum for small $p^\prime$ 
values. It then starts to resemble the cross section for electron scattering (upper row plots in Fig.~\ref{fig:tail}). 
Thus, for sufficiently small values of the lepton mass, the cross section has a distinct local maximum in the cross section 
in the small $p^\prime$ region. 
%

\begin{figure}[tbp]
\centering
         \includegraphics[scale=0.287]{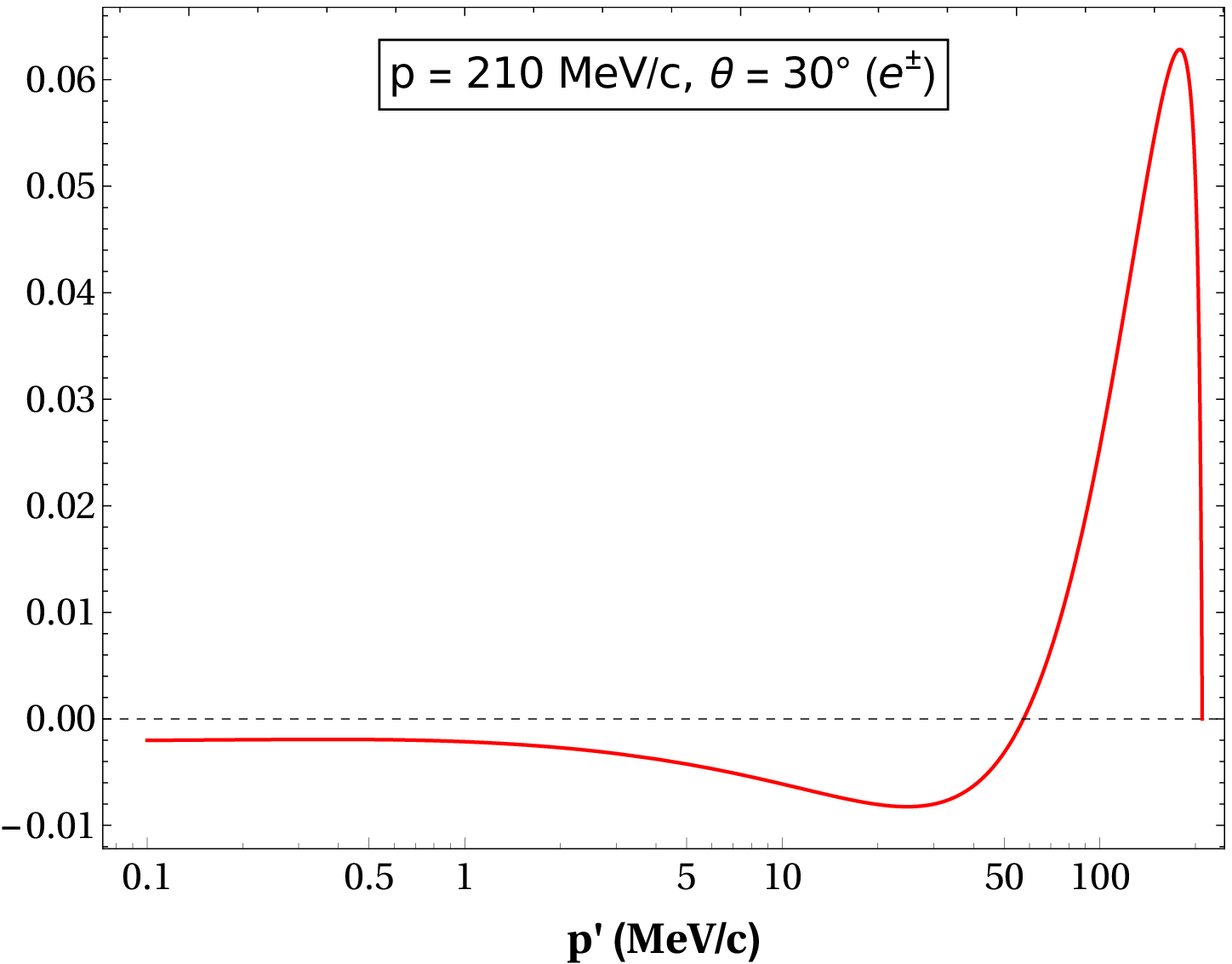} \includegraphics[scale=0.278]{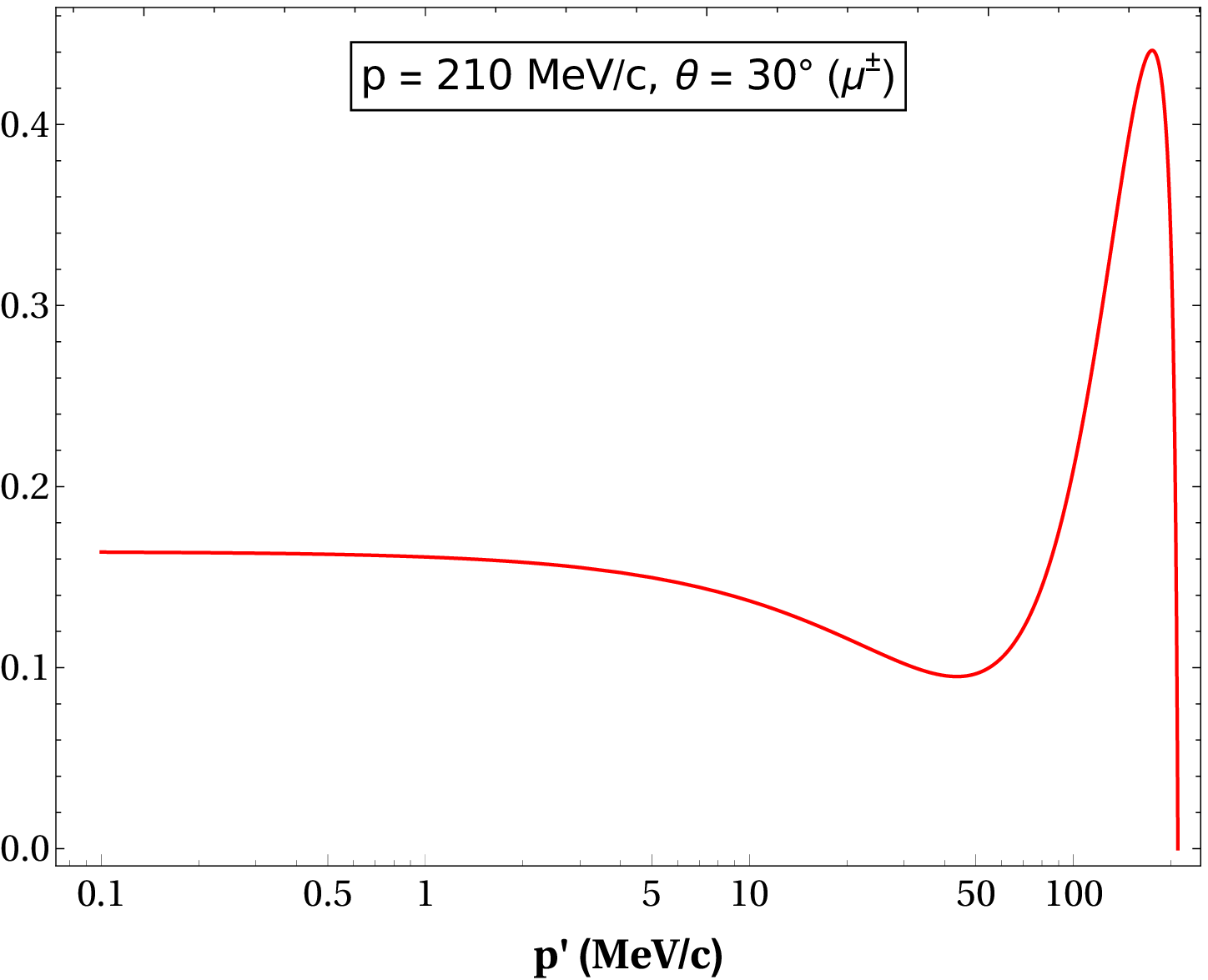}
\caption{The quantity $\delta_{\rm NLO}$ of Eq.~\eqref{eq:delta_NLO}, which stands for the NLO corrections 
relative to LO radiative tail cross section, is plotted as a function of the outgoing lepton momentum 
$p^\prime< p^\prime_{elastic}$, where $p^\prime_{elastic}\approx 203.8$ MeV/c for electron and about $203.1$ 
MeV/c for muon, for scattering angle $\theta=30\degree$. The left plot is for the electron case and the 
right plot gives the corresponding muon results. }
\label{fig:NLO_delta plot}
\end{figure}

As evident from Fig.~\ref{fig:tail}, our results up to and including the NLO appear to be dominated by the LO 
results. In other words, the  qualitative difference made by the NLO corrections to the electron and the muon cross 
sections are not at all apparent from the figure. Thus, for a better qualitative estimate of the NLO part of tail 
spectrum, we present a relative comparison between the LO and NLO contributions in terms of a quantity, 
$\delta_{\rm NLO}$, which measures the ratio of the NLO correction with respect to the LO contribution, namely  
\beq
\delta_{\rm NLO}=\left[\left(\frac{{\rm d}^2 \sigma^{\rm (NLO)}}{{\rm d}p^{\prime} {\rm d}\Omega_l}\right)-
\left(\frac{{\rm d}^2 \sigma^{\rm (LO)}}{{\rm d}p^{\prime} {\rm d}\Omega_l}\right)\right] \text{\huge /}
\left(\frac{{\rm d}^2 \sigma^{\rm (LO)}}{{\rm d}p^{\prime} {\rm d}\Omega_l}\right)\,. 
\label{eq:delta_NLO}
\eeq
This quantity is plotted in Fig.~\ref{fig:NLO_delta plot} against the outgoing lepton momentum $p^\prime$, corresponding 
to the same kinematics specification as in the right column plots in Fig.~\ref{fig:tail}. Clearly, the $\delta_{\rm NLO}$ 
for the muon spectrum is an order of magnitude larger than same for the electron spectrum. Moreover, in the 
electron case, the NLO corrections in the low $p^\prime$ region below 50 MeV/c become negative, in contrast to the NLO 
corrections for the muon that remain positive definite for the entire range of $p^\prime$ values. 
%

Finally, in comparing our results with the expression Eq.~(B.5) in Ref.~\cite{motsai1969}, we find that the expression 
has to be corrected as follows. The very first energy factor, $E_p / E_s$, multiplying the integral in that expression 
should have been $\vec{p}^{\,\,2} / (E_p E_s)$. This reduces to the energy factor in Eq.~(B.5) provided we neglect the 
lepton mass. Incorporating this correction and by ignoring the proton form factors (anomalous magnetic moment contribution 
also excluded) in  Eq.~(B.5) of Ref.~\cite{motsai1969}, we find  only a nominal difference with our LO result obtained 
from Eq.~(\ref{eq:sigma5}). However, once we include the full NLO result obtained by evaluating Eqs.~(\ref{eq:sigma6}) 
and (\ref{eq:sigma7}), which includes the $m_p^{-1}$ recoil corrections from the phase space, the $\delta$-function, and 
the matrix elements, the differences with Ref.~\cite{motsai1969} indeed becomes negligible, as evident from the right 
column plots in Figs.~\ref{fig:tail}. Note that Ref.~\cite{motsai1969} treated both the leptons and the proton in a 
common relativistic framework, whereas in our $\chi$PT approach the proton, being a heavy particle, is treated 
non-relativistically in both the phase-space expressions and the matrix elements. 

\section{Discussions and Summary} 
\label{section:summary}

\begin{figure}[t]
\centering
         \includegraphics[scale=0.287]{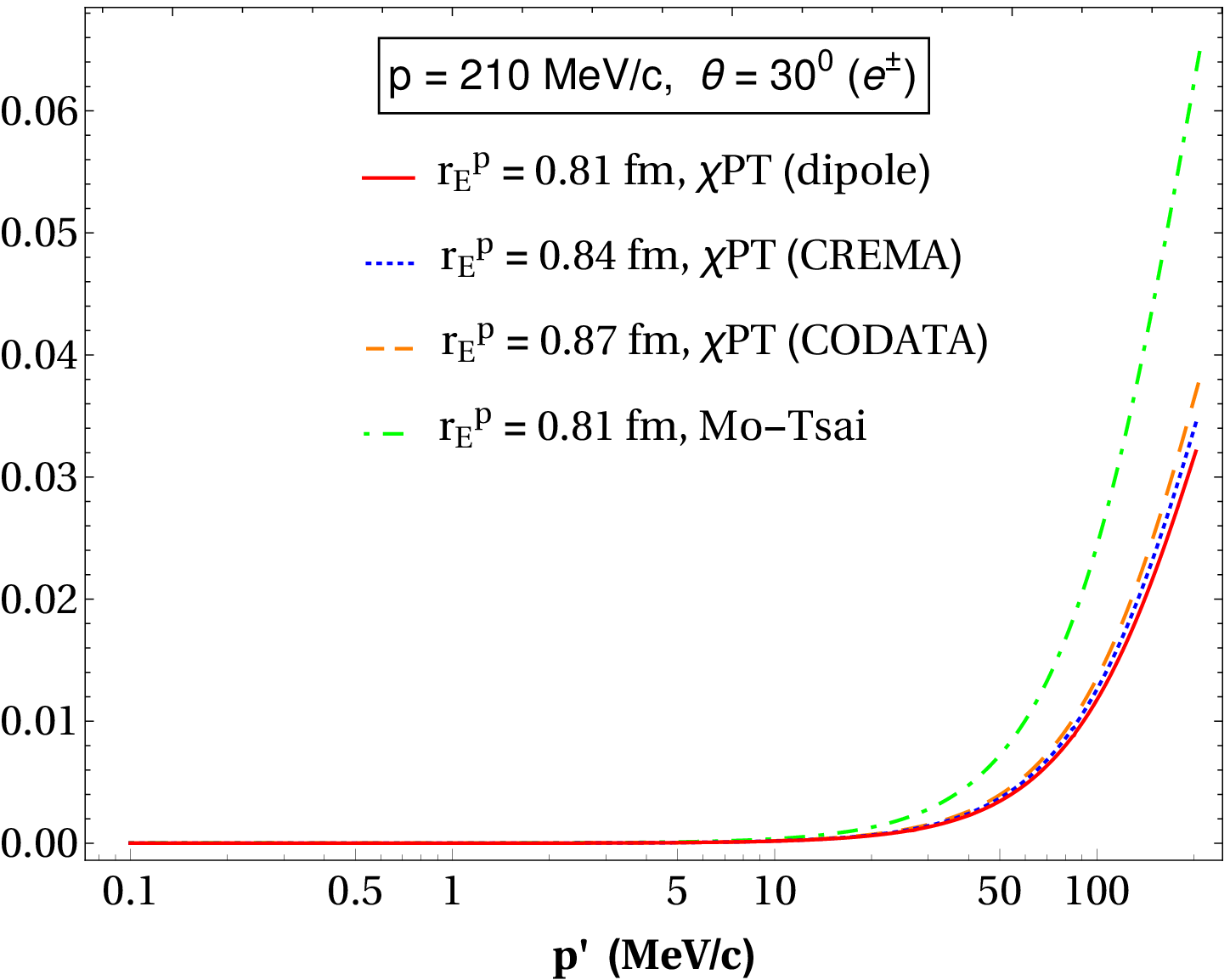} \includegraphics[scale=0.287]{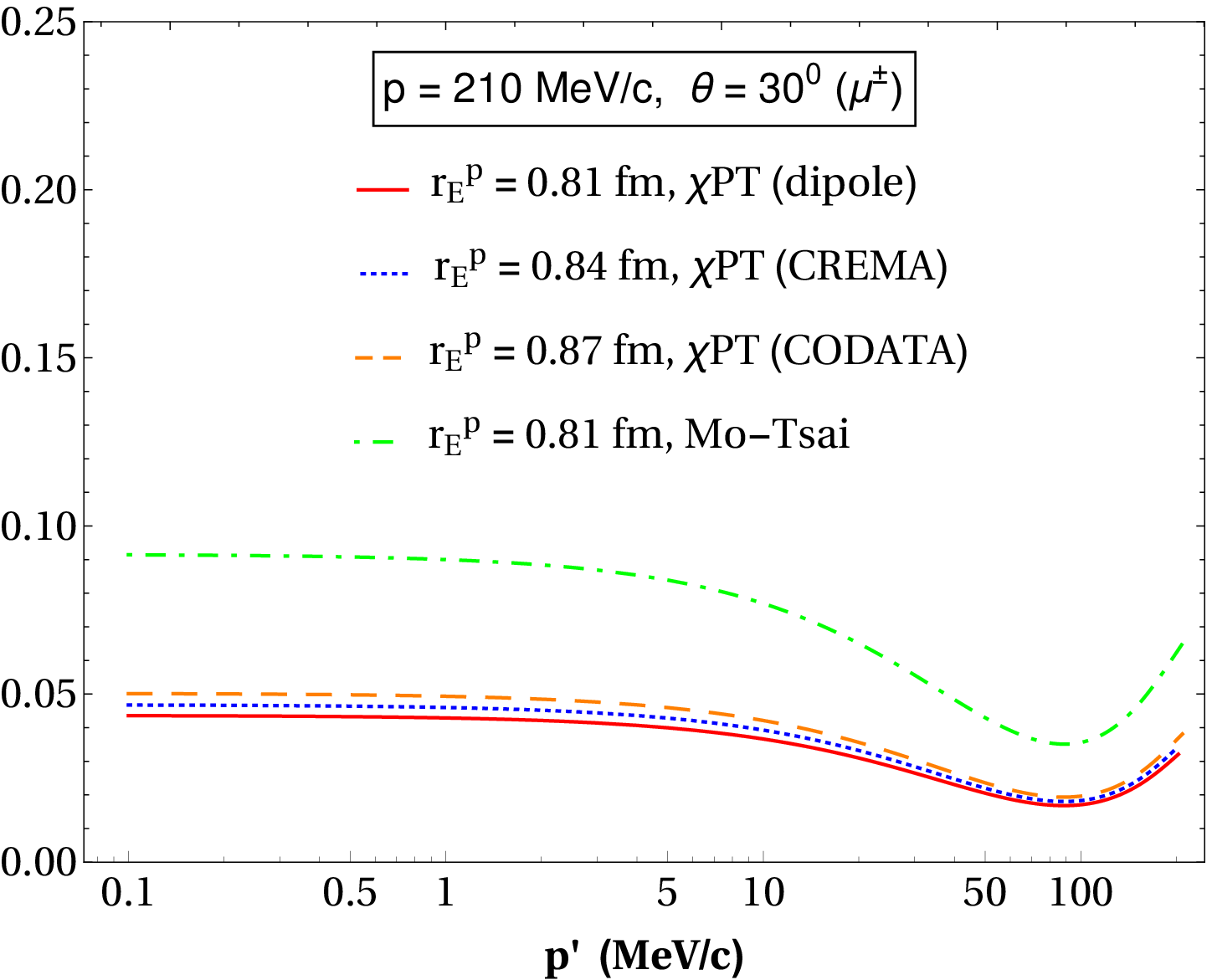}
\caption{The quantity, $|\delta_{\rm NNLO}|$ is a qualitative measure of the expected NNLO proton's r.m.s. radius 
contribution to the radiative tail cross section when the lepton is radiating. This quantity is plotted as a function of 
the outgoing lepton momentum $p^\prime< p^\prime_{elastic}$, where $p^\prime_{elastic}\approx 203.8$ MeV/c for electron and about 
$203.1$ MeV/c for muon, for scattering angle $\theta=30\degree$. Here different r.m.s. radii are used as phenomenological 
input to parametrize the ``$\chi$PT'' electric form factor $G^p_E$ (see text). For comparison, we also display the 
corresponding result obtained by using Eq.~(B.5) of Ref.~\cite{motsai1969}, while ignoring the magnetic form factor 
contribution. The left (right) plot corresponds to the electron (muon) results. }
\label{fig:structure plot}
\end{figure}

As mentioned in the introduction and in Sec.~$\rm II$, the pion loops, as well as $m_p^{-2}$ terms from 
${\cal L}_{\pi N}^{(2)}$  contribute at NNLO in $\chi$PT and generate the low-momentum  proton form factors. 
These contributions are supposed to be small according to standard $\chi$PT counting. However, as presented 
in Sec.~\ref{section:EFT} the {$\rm \chi PT$} counting scheme might not capture the fact that the probability 
of an electron radiating is much enhanced compared to the radiation from the heavy proton. Hence, we next examine 
the importance of the proton form factors generated by NNLO terms. The low-momentum NNLO expressions for the Sachs 
form factors $G^p_{E,M}$, have already been evaluated in earlier $\chi$PT works, e.g., Ref.~\cite{Bernard1998}, and 
they are effectively incorporated in our evaluation. We consider only those LO Feynman graphs where the real photon 
radiation originates from one of the lepton lines, i.e., diagrams (A) and (B) in Fig.~\ref{fig:LOfeyndiag}. We only 
include the dominant Sachs electric form factor $G^p_E$ of the proton, in particular the Taylor expanded form given by 
Eq.~\eqref{eq:GEM}, at the exchanged photon-proton vertices associated with the LO diagrams. In order to have a rough 
assessment of the relative importance of the proton's structure effects, we show our results as a ratio ${\mathscr R}$, 
which may be taken as a qualitative measure of the expected NNLO corrections $\delta_{\rm NNLO}$: 
\beq
{\mathscr{R}}=\frac{\left(\frac{{\rm d}^2 \sigma}{{\rm d}p^{\prime} {\rm d}\Omega^\prime_l}\right)_{\rm form} \!\!\! - \,\left(\frac{{\rm d}^2 \sigma}{{\rm d}p^{\prime} {\rm d}\Omega^\prime_l}\right)_{\rm point}}{\left(\frac{{\rm d}^2 \sigma}{{\rm d}p^{\prime} {\rm d}\Omega^\prime_l}\right)_{\rm point}}\,. 
\label{eq:delta_NNLO}
\eeq
Here, the subscripts ``form'' and ``point'', respectively, denotes the radiative tail differential cross section 
evaluated with and without the proton's r.m.s. charge radius $r^p_{E}$ included. Thus, in our analysis we approximate 
the NNLO $\chi$PT contributions as
\bea
\left(\frac{{\rm d}^2 \sigma}{{\rm d}p^{\prime}\, {\rm d}\Omega^\prime_l}\right)_{\rm form} \!\!\!\! &\to & \int^1_{-1}{\rm d}(\cos\alpha) \left(\frac{{\rm d}^3 \sigma^{\rm (LO)}}{{\rm d}p^{\prime}\, {\rm d}\Omega^\prime_l\, {\rm d}\cos\alpha}\right) G^p_E(q^2) \,,\nonumber \\
\left(\frac{{\rm d}^2 \sigma}{{\rm d}p^{\prime}\, {\rm d}\Omega^\prime_l}\right)_{\rm point} \!\!\!\! &\to & \frac{{\rm d}^2 \sigma^{\rm (LO)}}{{\rm d}p^{\prime}\, {\rm d}\Omega^\prime_l}\,,
\eea
with $q^2$ given by Eq.~\eqref{eq:q_squared}, and the charge radius $r^p_{E}$ of the proton to be used as the input. In 
phenomenological analyses, the {\it dipole} parametrization~\cite{dipole} is a commonly employed parametrization for the 
Sachs form factors, as was used in Ref.~\cite{motsai1969} that correspond to $r^p_{E}=0.81$~fm. Furthermore, the recent 
high-precision measurements from the study of muonic hydrogen spectroscopy by the CREMA experimental 
collaboration~\cite{pohl10,pohl13} led to the controversial value of $r^p_{E}=0.84$~fm, a result that is $\sim 7\sigma$ 
smaller than the previously accepted CODATA result, $r^p_{E}=0.87$~fm~\cite{Mohr2012}. These three values of $r^p_{E}$ 
are used as the ``measured'' r.m.s. radius input to $G^p_E$ in the above relation. The results are displayed in 
Fig.~\ref{fig:structure plot} where they are labelled as ``$\chi$PT (dople)'', ``$\chi$PT (CREMA)'' and ``$\chi$PT (CODATA)'', 
respectively. Finally, these NNLO predictions are compared with the corresponding result obtained by using Eq.~(B.5) of 
Ref.~\cite{motsai1969} (with our corrected version), with and without their phenomenological electric form factor $G^p_E$ 
for obtaining the ``form'' and ``point'' contributions, respectively, while ignoring their magnetic dipole form factor 
$G^p_M$. This result is labelled as ``Mo-Tsai'' in the figure. 

By comparing the plots in Figs.~\ref{fig:NLO_delta plot} and \ref{fig:structure plot}, we observe 
that the NLO and NNLO corrections relative to the LO results can be significant. As evident from the plots, 
for an incoming electron of momentum $p=210$ MeV/c, outgoing momentum $p^\prime =100$ MeV/c, and scattering angle 
$\theta = 30\degree$, we find that the NLO and NNLO corrections modify the LO radiative tail cross section by about 
$2\%$ and $1\%$, respectively, with the ``$\chi$PT''  form factors [i.e., Eq.~\eqref{eq:GEM} with the above 
phenomenological r.m.s. radii.] Using the same kinematics for the muon, the corresponding LO results get modified 
about $20\%$ and $2\%$, respectively. In addition, our NNLO results suggest that effectively the pion loops strongly 
suppress the local maximum at small $p'$ values in the electron radiative tail cross section. In case of the electron 
the crucial observation is that the NNLO form factor effects are of the same order as our NLO contributions. However, 
for the muon we find that the NNLO ``$\chi$PT'' contributions are about a factor of two smaller than the NLO 
contributions and more consistent with the standard $\chi$PT counting rules. But again in contrast, 
Fig.\ref{fig:structure plot} shows that for the muon ``Mo\,-Tsai'' result the proton's structure contributions are a 
factors of two larger than those of the NNLO ``$\chi$PT'' results, and therefore, of similar magnitude as that of the 
muon NLO result shown in Fig.~\ref{fig:NLO_delta plot}.  

In Fig.~\ref{fig:tail}, the bremsstrahlung cross sections diverge as the maximal value  $p^\prime_{max}$ approaches 
the elastic lepton-proton scattering value $p^\prime_{elastic}$. As mentioned this is due to the infrared divergence (IR) 
of the bremsstrahlung process when  the bremsstrahlung photon energy tends to zero. As demonstrated in, e.g.,  
Refs.~\cite{maximon00,motsai1969}, the cross section is free of IR singularities, provided the {\it virtual} radiative 
corrections are included in the calculation. In $\chi$PT, one can systematically evaluate the effect of virtual photon-loops, 
as well as the bremsstrahlung contribution from the leptons and protons in order to remove the IR singularities from 
observables. For the elastic lepton-proton scattering, the virtual photon loops along with the so-called 
{\it two-photon exchange} (TPE) contributions will additionally introduce ultraviolet (UV) divergences. These divergences 
can be treated systematically in $\chi$PT using the procedure of dimensional regularization which ensure gauge invariance 
at every perturbative order. To the best of our knowledge, such an evaluation has not been pursued to date in the context 
of low-energy lepton-proton scattering. This would naturally involve the introduction of low-energy constants (LECs) in the 
$\chi$PT Lagrangian required for the purpose of renormalization. Fortunately, at the order of our interest, all such LECs 
are known and can be taken directly from earlier $\chi$PT works, e.g., Ref.~\cite{Ando}. Such a systematic evaluation of 
the radiative corrections is beyond the scope of the present discussion, and shall be pursued in a future work.

%
%
In retrospect, the purpose of this work was to present a qualitative but yet pedagogical evaluations of the LO 
and NLO contributions to the lepton-proton bremsstrahlung cross section in the context of a low-energy 
EFT framework. Here we summarize some of the essential aspects of this work:
\begin{itemize}
\item
One important issue was to discuss a scenario where a large change in the angular spectrum of the bremsstrahlung 
process can be expected at typical momenta not much larger than the muon mass. The importance of such a study is 
very relevant to the MUSE experimental program where high precision lepton-proton scatterings at very low-momentum 
transfers will be pursued in order to investigate the reason behind the unexpectedly large discrepancy of the proton 
charge radius extracted from scattering experiments and the radius obtained from muonic hydrogen measurements. Our 
analysis demonstrates that a non-standard treatment of the bremsstrahlung corrections for muon scattering must be 
carefully thought through by the MUSE collaboration. 
\item 
In $\chi$PT employing Coulomb gauge, the two LO diagrams (C) and (D) in Fig.~~\ref{fig:LOfeyndiag}, and the two NLO 
diagrams, (I) and (J) in Fig.~\ref{fig:NLOfeyndiag}, do not contribute. In other words, bremsstrahlung radiation from a LO 
proton-photon vertex does not contribute in our $\chi$PT analysis. However, there is non-trivial proton bremsstrahlung 
contributions [diagrams (G) and (H) in Fig.~\ref{fig:NLOfeyndiag}] at NLO associated with chiral order $\nu=1$ proton-photon 
vertex in $\chi$PT.  
\item
While taking the trace over the amplitude squared in order to determine the unpolarized bremsstrahlung cross section, 
the spin dependent interactions in the NLO Lagrangian give vanishing contribution. Consequently, the anomalous magnetic 
moment of the proton does not contribute to the cross section for this process at NLO in $\chi$PT, contrary 
to usual expectations. Thus, the proton's magnetic moment starts contributing to the bremsstrahlung process only 
at NNLO, as implicitly parametrized in our case in terms of the proton's r.m.s. radius. 
\item
Our results displayed in Figs.~\ref{fig:peak_e} and \ref{fig:peak_mu}, clearly indicate that the widely used 
peaking approximation in radiative analysis of high energy electron scattering data can not be used in the 
radiative analyses of the future low-energy muon data from MUSE.
\item
As evidenced from Fig.~\ref{fig:tail}, the radiative tail cross section for the electron scattering process for 
small outgoing momenta, $p^\prime\lesssim 5$ MeV/c exhibits a local maximum that can be attributed due to the small 
but non-zero electron mass. However, in the same figure we show that such a behavior disappears for the much larger 
lepton muon mass. Since MUSE is designed only to detect the lepton scattering angle $\theta$ and can not determine 
the value of the outgoing lepton momentum $|\vec{p}^{\,\prime}\,|$, care must be taken in the analysis of the MUSE 
data when one corrects for the radiative tail from the bremsstrahlung spectra. 
\item
At NNLO the proton's r.m.s. radius enters the evaluation via the LECs in ${\cal L}_{\pi N}^{(2)}$, meaning that the 
low-momentum aspects of the proton's structure (r.m.s. radius) naturally contribute at this sub-leading order. In 
particular, we find that NNLO corrections for electron scattering bremsstrahlung diagrams illustrated in 
Fig.~\ref{fig:NNLOfeyndiag}, are as large as the NLO corrections, contrary to what is expected in standard chiral 
power counting. 
\item
Finally, we report that our evaluations revealed a misprint in the overall energy factor in the well-known review 
article~\cite{motsai1969}. 
\end{itemize}
As a future extension of this work, we intend to improve on the systematic uncertainties and evaluate the radiative 
corrections to the lepton-proton bremsstrahlung scattering process (including the two-photon exchange corrections) 
which should eliminate the unphysical IR singularities in the large $p^\prime$ region in all the differential cross 
sections considered in this work.   

\section{Acknowledgments}
Numerous discussions with Steffen Strauch have been very helpful in the present work. We thank 
Ulf Mei{\ss}ner, Daniel Phillips and Shung-Ichi Ando for useful comments. One of us (FM) is grateful for the 
hospitality at Ruhr University Bochum while working on this project. 

\appendix

\section{The partial NLO  amplitudes in Eq.~(\ref{eq:sigma7}) }
\label{sec:AppendixA}
We display the eight partial amplitudes appearing in our NLO corrections to the lepton-proton bremsstrahlung 
differential cross section, Eq.~\eqref{eq:sigma7}, namely
\begin{widetext}
\bea
W_{AE} &=&-\, m_l^2E_l (\vec{p}_p^{\,\prime}\cdot\vec{p}^{\,\prime}) - m_l^2E_l (\vec{p}_p\cdot\vec{p}^{\,\prime}) + m_l^2E^\prime_l (\vec{p}_p^{\,\prime}\cdot\vec{k}) + m_l^2E^\prime_l (\vec{p}_p\cdot\vec{k}) - m_l^2E^\prime_l (\vec{p}_p^{\,\prime}\cdot\vec{p}) - m_l^2E^\prime_l (\vec{p}_p\cdot\vec{p}) + m_l^2E^0_{\gamma} (\vec{p}_p^{\,\prime}\cdot\vec{p}^{\,\prime})
\nonumber\\
&& +\, m_l^2E^0_{\gamma}(\vec{p}_p\cdot\vec{p}^{\,\prime}) + E_l(E^0_{\gamma})^2 (\vec{p}_p\cdot\vec{p}^{\,\prime}) + E_l(E^0_{\gamma})^2 (\vec{p}_p^{\,\prime}\cdot\vec{p}^{\,\prime}) + E_lE^\prime_lE^0_{\gamma} (\vec{p}_p\cdot\vec{k}) + E_lE^\prime_lE^0_{\gamma} (\vec{p}_p^{\,\prime}\cdot\vec{k}) - E^\prime_l (\vec{p}_p^{\,\prime}\cdot\vec{k})(\vec{p}\cdot\vec{k})
\nonumber\\
&& -\, E^\prime_l (\vec{p}_p\cdot\vec{k})(\vec{p}\cdot\vec{k}) - E^0_{\gamma} (\vec{p}_p^{\,\prime}\cdot\vec{p}^{\,\prime})(\vec{p}\cdot\vec{k}) - E^0_{\gamma} (\vec{p}_p\cdot\vec{p}^{\,\prime})(\vec{p}\cdot\vec{k}) \,,
\nonumber\\  \nonumber\\
W_{AF}&=&m_l^2E_l (\vec{p}_p^{\,\prime}\cdot\vec{k})  + m_l^2E_l (\vec{p}_p\cdot\vec{k}) + m_l^2E^\prime_l (\vec{p}_p^{\,\prime}\cdot\vec{k}) + m_l^2E^\prime_l (\vec{p}_p\cdot\vec{k}) + m_l^2E^0_{\gamma} (\vec{p}_p^{\,\prime}\cdot\vec{k}) + m_l^2E^0_{\gamma} (\vec{p}_p\cdot\vec{p}) + m_l^2E^0_{\gamma} (\vec{p}_p\cdot\vec{p}^{\,\prime}) 
\nonumber\\
&& +\, m_l^2E^0_{\gamma}(\vec{p}_p\cdot\vec{k}) - m_l^2E^0_{\gamma} (\vec{p}_p^{\,\prime}\cdot\vec{p}^{\,\prime}) - m_l^2E^0_{\gamma} (\vec{p}_p^{\,\prime}\cdot\vec{p}) + E_l^2E^\prime_l (\vec{p}_p^{\,\prime}\cdot\vec{p}^{\,\prime}) + E_l^2E^\prime_l (\vec{p}_p\cdot\vec{p}^{\,\prime}) + E_l^2E^0_{\gamma}(\vec{p}_p^{\,\prime}\cdot\vec{p}^{\,\prime}) 
\nonumber\\
&& +\, E_l^2E^0_{\gamma} (\vec{p}_p\cdot\vec{p}^{\,\prime}) + E_l(E^\prime_l)^2 (\vec{p}_p^{\,\prime}\cdot\vec{p}) + E_l(E^\prime_l)^2 (\vec{p}_p\cdot\vec{p}) - E_l (E^\prime_l)^2 (\vec{p}_p^{\,\prime}\cdot\vec{k}) - E_l(E^\prime_l)^2 (\vec{p}_p\cdot\vec{k}) + E_l (\vec{p}_p^{\,\prime}\cdot\vec{p})(\vec{p}^{\,\prime}\cdot\vec{k}) 
\nonumber\\
&& +\, E_l (\vec{p}_p^{\,\prime}\cdot\vec{p}^{\,\prime})(\vec{p}^{\,\prime}\cdot\vec{k}) + E_l(\vec{p}_p\cdot\vec{p})(\vec{p}^{\,\prime}\cdot\vec{k}) + E_l (\vec{p}_p\cdot\vec{p}^{\,\prime})(\vec{p}^{\,\prime}\cdot\vec{k}) - E_l (\vec{p}_p^{\,\prime}\cdot\vec{p}^{\,\prime})(\vec{p}\cdot\vec{p}^{\,\prime}) - E_l (\vec{p}_p^{\,\prime}\cdot\vec{p}^{\,\prime})(\vec{p}\cdot\vec{k}) 
\nonumber\\
&& -\, E_l (\vec{p}_p\cdot\vec{p}^{\,\prime})(\vec{p}\cdot\vec{p}^{\,\prime}) - E_l (\vec{p}_p\cdot\vec{p}^{\,\prime})(\vec{p}\cdot\vec{k}) + E^\prime_l (\vec{p}_p^{\,\prime}\cdot\vec{k})(\vec{p}\cdot\vec{p}^{\,\prime}) + E^\prime_l (\vec{p}_p\cdot\vec{k})(\vec{p}\cdot\vec{p}^{\,\prime}) - E^\prime_l (\vec{p}_p\cdot\vec{p})(\vec{p}\cdot\vec{p}^{\,\prime}) 
\nonumber\\
&& -\, E^\prime_l (\vec{p}_p\cdot\vec{p}^{\,\prime})(\vec{ p}\cdot\vec{k}) - E^\prime_l (\vec{p}_p^{\,\prime}\cdot\vec{p})(\vec{p}\cdot\vec{p}^{\,\prime}) - E^\prime_l (\vec{p}_p^{\,\prime}\cdot\vec{p}^{\,\prime})(\vec{p}\cdot\vec{k}) - E^0_{\gamma}(\vec{p}_p^{\,\prime}\cdot\vec{p})(\vec{p}\cdot\vec{p}^{\,\prime}) + E^0_{\gamma} (\vec{p}_p\cdot\vec{p})(\vec{p}\cdot\vec{p}^{\,\prime})  \,,
\nonumber\\  \nonumber\\ 
W_{AG}&=& m_l^2 E_l^2  E^0_{\gamma} + E_l^3E^{\prime}_lE^0_{\gamma} + 2m_l^2E_l (\vec{p}_p\cdot\vec{p}) - m_l^2 E_l (\vec{p}_p\cdot\vec{k}) - m_l^2E_l (\vec{p}\cdot\vec{k}) + 2E_l^2E^{\prime}_l (\vec{p}_p\cdot\vec{p}) - E_l^2 E^{\prime}_l(\vec{p}_p\cdot\vec{k}) 
\nonumber\\
&& -\, E_l^2 E^{\prime}_l(\vec{p}\cdot\vec{k}) - E^{\prime}_l E_l E^0_{\gamma} (\vec{p}_p\cdot\vec{p}) + E_l^2 E^0_{\gamma} (\vec{p}\cdot\vec{p}^{\,\prime}) - E_l^2 E^0_{\gamma} (\vec{p}_p\cdot\vec{p}^{\,\prime}) + 2E_l (\vec{p}_p\cdot\vec{p})(\vec{p}\cdot\vec{p}^{\,\prime}) - E_l (\vec{p}_p\cdot\vec{p})(\vec{p}^{\,\prime}\cdot\vec{k}) 
\nonumber\\
&& +\, E_l (\vec{p}_p\cdot\vec{p}^{\,\prime})(\vec{p}\cdot\vec{k}) - E_l (\vec{p}_p\cdot\vec{k})(\vec{p}\cdot\vec{p}^{\,\prime}) - E_l (\vec{p}\cdot\vec{p}^{\,\prime})(\vec{p}\cdot\vec{k}) \,,
\nonumber\\  \nonumber\\
W_{AH}&=&m_l^2E_l^2 E^0_{\gamma} + E_l^3 E_l^\prime E^0_{\gamma} + m_l^2 E_l (\vec{p}_p^{\,\prime}\cdot\vec{k}) - 2m_l^2E_l (\vec{p}_p^{\,\prime}\cdot\vec{p}) - m_l^2 E_l (\vec{p}\cdot\vec{k}) - 2E_l^2 E^{\prime}_l (\vec{p}_p^{\,\prime}\cdot\vec{p}) + E_l^2 E^{\prime}_l (\vec{p}_p^{\,\prime}\cdot\vec{k}) 
\nonumber\\
 && +\, E^{\prime}_l E_l E^0_{\gamma} (\vec{p}_p^{\, \prime}\cdot\vec{p}) - E_l^2 E^{\prime}_l(\vec{p}\cdot\vec{k}) + E_l^2 E^0_{\gamma} (\vec{p}_p^{\,\prime}\cdot\vec{p}^{\,\prime}) + E_l^2 E^0_{\gamma} (\vec{p}\cdot\vec{p}^{\,\prime}) + E_l (\vec{p}_p^{\,\prime}\cdot\vec{p})(\vec{p}^{\,\prime}\cdot\vec{k}) - 2E_l (\vec{p}_p^{\,\prime}\cdot\vec{p})(\vec{p}^{\,\prime}\cdot\vec{p})  
\nonumber\\
 && +\, E_l (\vec{p}_p^{\,\prime}\cdot\vec{k})(\vec{p}^{\,\prime}\cdot\vec{p}) - E_l (\vec{p}_p^{\,\prime}\cdot\vec{p}^{\,\prime})(\vec{p}\cdot\vec{k}) - E_l (\vec{p}^{\,\prime}\cdot\vec{p})(\vec{p}\cdot\vec{k}) \,.
\nonumber\\
W_{BE} &=&-\, m_l^2E_l (\vec{p}_p\cdot\vec{k}) - m_l^2 E_l (\vec{p}_p^{\,\prime}\cdot \vec{k}) - m_l^2 E_l^\prime (\vec{p}_p^{\,\prime}\cdot\vec{k}) - m_l^2E_l^\prime (\vec{p}_p\cdot\vec{k}) + m_l^2E^0_\gamma (\vec{p}_p\cdot\vec{p}^{\,\prime}) + m_l^2E^0_\gamma (\vec{p}_p^{\,\prime}\cdot\vec{p}^{\,\prime}) + m_l^2E^0_\gamma (\vec{p}_p\cdot\vec{p})
\nonumber\\
&& -\, m_l^2E^0_\gamma (\vec{p}_p^{\,\prime}\cdot\vec{k}) - m_l^2E^0_\gamma (\vec{p}_p^{\,\prime}\cdot\vec{p}) - m_l^2E^0_\gamma (\vec{p}_p\cdot\vec{k}) + E_l^2E^\prime_l (\vec{p}_p^{\,\prime}\cdot\vec{p}^{\,\prime}) +  E_l^2E^\prime_l (\vec{p}_p\cdot\vec{p}^{\,\prime}) +  E_l^2E^\prime_l (\vec{p}_p\cdot\vec{k}) +  E_l^2 E^\prime_l (\vec{p}_p^{\,\prime}\cdot\vec{k})
\nonumber\\
&&  +\, E_l (E_l^\prime)^2 (\vec{p}_p^{\,\prime}\cdot\vec{p}) + E_l (E_l^\prime)^2 (\vec{p}_p\cdot\vec{p}) - (E_l^\prime)^2 E^0_\gamma (\vec{p}_p^{\,\prime}\cdot\vec{p}) - (E_l^\prime)^2 E^0_\gamma (\vec{p}_p\cdot\vec{p}) + E_l (\vec{p}_p^{\,\prime}\cdot\vec{p})(\vec{p}^{\,\prime}\cdot\vec{k}) + E_l (\vec{p}_p\cdot\vec{p})(\vec{p}^{\,\prime}\cdot\vec{k}) 
\nonumber\\
&& -\, E_l (\vec{p}_p\cdot\vec{k})(\vec{p}\cdot\vec{p}^{\,\prime}) - E_l (\vec{p}_p^{\,\prime}\cdot\vec{p}^{\,\prime})(\vec{p}^{\,\prime}\cdot\vec{p}) - E_l (\vec{p}_p^{\,\prime}\cdot\vec{k})(\vec{p}^{\,\prime}\cdot\vec{p}) - E_l (\vec{p}_p\cdot\vec{p}^{\,\prime})(\vec{p}^{\,\prime}\cdot\vec{p}) + E_l^\prime (\vec{p}_p^{\,\prime}\cdot\vec{p})(\vec{p}^{\,\prime}\cdot\vec{k}) 
\nonumber\\
&& +\, E_l^\prime(\vec{p}_p\cdot\vec{p})(\vec{p}^{\,\prime}\cdot\vec{k}) - E_l^\prime (\vec{p}_p^{\,\prime}\cdot\vec{p})(\vec{p}^{\,\prime}\cdot\vec{p}) - E_l^\prime (\vec{p}_p^{\,\prime}\cdot\vec{p})(\vec{p}\cdot\vec{k}) - E_l^\prime(\vec{p}_p^{\,\prime}\cdot\vec{p}^{\,\prime})(\vec{p}\cdot\vec{k}) - E_l^\prime(\vec{p}_p\cdot\vec{p})(\vec{p}\cdot\vec{p}^{\,\prime})
\nonumber\\
&& -\, E_l^\prime (\vec{p}_p\cdot\vec{p})(\vec{p}\cdot\vec{k}) - E_l^\prime (\vec{p}_p\cdot\vec{p}^{\,\prime})(\vec{p}\cdot\vec{k}) + E^0_\gamma (\vec{p}_p^{\,\prime}\cdot\vec{p}^{\,\prime})(\vec{p}\cdot\vec{p}^{\,\prime}) + E^0_\gamma (\vec{p}_p\cdot\vec{p}^{\,\prime})(\vec{p}\cdot\vec{p}^{\,\prime}) \,,
\nonumber\\  \nonumber\\ 
W_{BF}&=& -\, m_l^2 E_l (\vec{p}_p^{\,\prime}\cdot\vec{p}^{\,\prime}) - m_l^2 E_l (\vec{p}_p\cdot\vec{p}^{\,\prime}) - m_l^2 E_l (\vec{p}_p^{\,\prime}\cdot \vec{k}) - m_l^2 E_l (\vec{p}_p\cdot\vec{k}) - m_l^2 E_l^\prime (\vec{p}_p^{\,\prime}\cdot\vec{p})- m_l^2 E_l^\prime (\vec{p}_p\cdot\vec{p}) - m_l^2 E^0_\gamma (\vec{p}_p^{\,\prime}\cdot\vec{p})  
\nonumber\\ 
&& - m_l^2 E^0_\gamma (\vec{p}_p\cdot\vec{p})+\, E_l^\prime (E^0_\gamma)^2 (\vec{p}_p^{\,\prime}\cdot\vec{p}) + E_l^\prime (E^0_\gamma)^2 (\vec{p}_p\cdot\vec{p}) + E_l E_l^\prime E^0_\gamma (\vec{p}_p^{\,\prime}\cdot\vec{k}) + E_l E_l^\prime E^0_\gamma (\vec{p}_p\cdot\vec{k}) - E_l (\vec{p}_p^{\,\prime}\cdot\vec{k})(\vec{p}^{\,\prime}\cdot\vec{k})   
\nonumber\\
&&- E_l (\vec{p}_p\cdot\vec{k})(\vec{p}^{\,\prime}\cdot\vec{k})- E^0_\gamma (\vec{p}_p^{\,\prime}\cdot\vec{p})(\vec{p}^{\,\prime}\cdot\vec{k}) - E^0_\gamma (\vec{p}_p\cdot\vec{p})(\vec{p}^{\,\prime}\cdot\vec{k}) \,,
\nonumber\\  \nonumber\\
W_{BG} &=&m_l^2 (E_l^\prime)^2 E^0_\gamma +  E_l E^0_\gamma (E_l^\prime)^3 + 2m_l^2 E_l^\prime (\vec{p}_p\cdot\vec{p}^{\,\prime}) + m_l^2 E_l^\prime (\vec{p}_p\cdot\vec{k}) - m_l^2 E_l^\prime (\vec{p}^{\,\,\prime}\cdot\vec{k}) + 2 E_l (E_l^\prime)^2 (\vec{p}_p\cdot\vec{p}^{\,\prime}) + E_l (E_l^\prime)^2 (\vec{p}_p\cdot\vec{k}) 
\nonumber\\
&& -\, E_l (E_l^\prime)^2 (\vec{p}^{\,\prime}\cdot\vec{k}) + E_l E_l^\prime E^0_\gamma (\vec{p}_p\cdot\vec{p}^{\,\prime}) + (E_l^\prime)^2 E^0_\gamma (\vec{p}_p\cdot\vec{p}) +(E_l^\prime)^2 E^0_\gamma (\vec{p}^{\,\prime}\cdot\vec{p}) + 2E_l^\prime (\vec{p}_p\cdot\vec{p}^{\,\prime})(\vec{p}^{\,\prime}\cdot\vec{p}) + E_l^\prime (\vec{p}_p\cdot\vec{k})(\vec{p}\cdot\vec{p}^{\,\prime}) 
\nonumber\\
&& +\, E_l^\prime (\vec{p}_p\cdot\vec{p}^{\,\prime})(\vec{p}\cdot\vec{k}) - E_l^\prime (\vec{p}^{\,\prime}\cdot\vec{k})(\vec{p}^{\,\prime}\cdot\vec{p}) - E_l^\prime (\vec{p}_p\cdot \vec{p})(\vec{p}^{\,\prime}\cdot\vec{k}) \,, 
\nonumber\\  \nonumber\\ 
W_{BH} &=& -\, m_l^2 (E_l^\prime)^2 E^0_\gamma - E_l (E_l^\prime)^3E^0_\gamma + m_l^2 E_l^\prime (\vec{p}_p^{\,\prime}\cdot\vec{k}) + m_l^2 E_l^\prime (\vec{p}^{\,\prime}\cdot\vec{k}) + 2m_l^2 E_l^\prime (\vec{p}_p^{\,\prime}\cdot\vec{p}^{\,\prime}) + 2 E_l (E_l^\prime)^2 (\vec{p}_p^{\,\prime}\cdot\vec{p}^{\,\prime}) + E^2_l (E_l^\prime)^2 (\vec{p}_p^{\,\prime}\cdot\vec{k}) 
\nonumber\\
&&  +\, E_l (E_l^\prime)^2 (\vec{p}^{\,\prime}\cdot \vec{k}) + E_l E_l^\prime E^0_\gamma (\vec{p}_p^{\,\prime}\cdot\vec{p}^{\,\prime}) + (E_l^\prime)^2 E^0_\gamma (\vec{p}_p^{\,\prime}\cdot\vec{p}) - (E_l^\prime)^2 E^0_\gamma (\vec{p}^{\,\prime}\cdot\vec{p}) + 2 E_l^\prime (\vec{p}_p^{\,\prime}\cdot\vec{p}^{\,\prime})(\vec{p}^{\,\prime}\cdot\vec{p}) + E_l^\prime (\vec{p}_p^{\,\prime}\cdot\vec{p}^{\,\prime})(\vec{p}\cdot\vec{k})  
\nonumber\\
&&  +\, E_l^\prime (\vec{p}_p^{\,\prime}\cdot \vec{k})(\vec{p}^{\,\prime}\cdot\vec{p}) + E_l^\prime (\vec{p}^{\,\prime}\cdot\vec{k})(\vec{p}^{\,\prime}\cdot\vec{p}) - E_l^\prime (\vec{p}_p^{\,\prime}\cdot\vec{p})(\vec{p}^{\,\prime}\cdot\vec{k}) \,.
\nonumber\\
\eea
\end{widetext}
In the rest frame of hadron target, the three-momentum of the proton, $|\vec{p}\,|=0$. The 
expressions for the dot-products in our preferred frame of reference, Fig.~\ref{fig:kin_ref},  
are:  
\bea
\vec{p}\cdot\vec{p}^{\,\prime} &=& |\vec{p}^{\,\prime}| |\vec{p}\,| \cos\theta\,, 
\nonumber\\
\vec{k}\cdot\vec{p}^{\,\prime} &=& E_\gamma|\vec{p}^{\,\prime}| (\cos\gamma\,\cos\alpha + \sin\alpha\,\sin\gamma\,\cos\phi_\gamma)\,, 
\nonumber\\
\vec{k}\cdot\vec{p} &=& E_\gamma|\vec{p}\,| (\cos\zeta\,\cos\alpha + \sin\alpha\,\sin\zeta\,\cos\phi_\gamma)\,, 
\nonumber\\
\vec{p}_p^{\,\prime}\cdot\vec{p}^{\,\prime} &=& (\vec{Q}-\vec{k})\cdot\vec{p}^{\,\prime} = |\vec{Q}| |\vec{p}^{\,\prime}| \cos\gamma - \vec{k}\cdot\vec{p}^{\,\prime}\,,
\nonumber\\
\vec{p}_p^{\,\prime}\cdot\vec{p} &=& (\vec{Q}-\vec{k})\cdot\vec{p} = |\vec{Q}| |\vec{p}\,| \cos\zeta - \vec{k}\cdot\vec{p}\,,
\nonumber\\
\vec{p}_p^{\,\prime}\cdot\vec{k} &=& (\vec{Q}-\vec{k})\cdot\vec{k} = |\vec{Q}| E_\gamma \cos\alpha - E_\gamma^2\,.
\eea 


\end{document}